\documentstyle[multicol,prl,aps,psfig,amsfonts]{revtex}
\begin{document}
\psfigurepath{fig}

\title{The Ground State of a Two-Dimensional Electron Liquid
in a Weak Magnetic Field}
\author{M. M. Fogler, A. A. Koulakov, and B. I. Shklovskii}

\address{Theoretical Physics Institute, University of Minnesota,
116 Church St. Southeast, Minneapolis, Minnesota 55455}

\date{January 23, 1996}

\maketitle

\begin{abstract}

We study the ground state of a clean two-dimensional electron liquid in a
weak magnetic field where $N \gg 1$ lower Landau levels are completely filled
and the upper level is partially filled.  It is shown that the electrons at
the upper Landau level form domains with filling factor equal to one and
zero. The domains alternate with a spatial period of the order of the
cyclotron radius, which is much larger than the interparticle distance at the
upper Landau level. The one-particle density of states, which can be probed
by tunneling experiments, is shown to have a pseudogap linearly dependent on
the magnetic field in the limit of large $N$.

\end{abstract}

\begin{multicols}{2}
\section{Introduction}
\label{Introduction}

The ground state of an interacting two-dimensional (2D) electron gas in a
magnetic field has been the subject of a persistent attention. The studies
have been focused mostly on the case of very strong magnetic fields where
only the lowest Landau level (LL) is occupied, so that the filling factor
$\nu = k_{\rm F}^2 l^2$ does not exceed unity (here $k_{\rm F}$ is the Fermi
wave-vector of the 2D gas in zero magnetic field and $l$ is the magnetic
length, $l^2 = \hbar / m \omega_c$\/). The physics at the lowest LL turned
out to be so rich that, perhaps, only at $\nu = 1$ the ground state has a
simple structure.  Namely, it corresponds to one fully occupied spin subband
of the lowest LL. The charge density in such a state is uniform. The case of
a partial filling, $\nu < 1$, is much more interesting. Using the
Hartree-Fock (HF) approximation, Fukuyama {\em et al.}~\cite{Fukuyama} found
that a uniform uncorrelated spin-polarized electron liquid (UEL) is unstable
against the formation of a charge density wave (CDW). The CDW instability
occurs at wave-vectors $q > 0.79 l^{-1}$.  Later, Yoshioka
and Fukuyama and also Yoshioka and Lee~\cite{Yoshioka} claimed that the
optimal period of the CDW coincides with that of the classical Wigner crystal
(WC). The difference of their HF WC state from a classical WC of point-like
particles is that the electrons are smeared over a distance of order $l$
around the sites of the WC lattice.  The wave-function for this state was
written in a simple form by Maki and Zotos~\cite{Maki}.

Later, however, it turned out that non-HF trial states suggested by
Laughlin~\cite{Laughlin} for $\nu = \frac13, \frac15$ to explain the
fractional quantum Hall effect are by a few percent lower in energy. The
Laughlin states were further interpreted in terms of an integer number of
fully occupied LLs of new quasiparticles, composite fermions~\cite{Jain}.
This concept was then applied to even denominator fractions~\cite{HLR}.
Thus, although the HF approximation gives a rather accurate estimate of the
energy, it fails to describe electron-electron correlations at a partially
filled lowest LL.

Recently, the requirement of the complete spin polarization in the ground
state was also reconsidered. It was found that a partially filled lowest LL
may contain skyrmions~\cite{Sondhi}.

In this paper we consider the case of weak magnetic fields or high LL numbers
$N$. A short version of this work~\cite{Koulakov} was published before.

There is growing evidence from analytical and numerical calculations that
fractional states, composite fermions and skyrmions are restricted to the two
lowest LLs ($N = 0, 1$) only (see
Refs.~\onlinecite{Belkhir_Morf,Wu,Aleiner}).
This point of view is also consistent with the experiment because none of
those structures has been observed for $N > 1$. Denote by $\bar{\nu}_N$ the
filling of the upper LL, $\bar{\nu}_N = \nu - 2 N$. Based on the arguments
above, we will assume that (i) at $\bar{\nu}_N \leq 1$ the upper LL is
completely spin-polarized and (ii) the HF approximation gives an adequate
description of the system.

Our theory strongly relies on the existence of Landau levels. In other words,
we assume that even in weak magnetic fields, where the cyclotron gap
$\hbar\omega_c$ is small, the electron-electron interactions do not destroy
the Landau quantization. Certainly, this is far from being evident. To see
that the LL mixing is indeed small, one has to calculate the interaction
energy per particle at the upper LL and verify that its absolute value is
much smaller than $\hbar\omega_c$. The largest value of the interaction
energy is attained at $\bar{\nu}_N = 1$ where the electron density at the
upper LL is the largest. The interaction energy per particle is equal to
$-\frac12 E_{\rm ex}$, where $E_{\rm ex}$ is the exchange-enhanced gap for
the spin-flip excitations~\cite{Enhancement} at $\bar{\nu}_N = 1$ (it
determines, e.g., the activation energy between spin-resolved quantum Hall
resistivity peaks). Aleiner and Glazman~\cite{Aleiner} calculated
$E_{\rm ex}$ to be
\begin{equation}
E_{\rm ex} = \frac{r_s\hbar\omega_c}{\sqrt{2}\,\pi}
\ln\left(\frac{2\sqrt{2}}{r_s}\right) + E_{\rm h}, \quad r_s \ll 1,
\label{E_ex}
\end{equation}
where $E_{\rm h}$ is the hydrodynamic term (in terminology of
Ref.~\onlinecite{Aleiner}) given by~\cite{Comment on exchange}
\begin{equation}
          E_{\rm h} = \hbar\omega_c \frac{\ln(N r_s)}{2 N + 1}.
\label{E_h}
\end{equation}
The parameter $r_s$ entering these formulae is defined by $r_s = \sqrt{2} /
k_{\rm F}a_{\rm B}$, $a_{\rm B} = \hbar^2\kappa / m e^2$ being the effective
Bohr radius. Therefore, in the considered limit $r_s \ll 1$ the LLs are
indeed preserved. In realistic samples $r_s \sim 1$ but even at such $r_s$
the ratio $\alpha = E_{\rm ex} / \hbar\omega_c$ is still rather small.
Experimentally, this ratio can be estimated to be near $0.25$ at $0 \leq N
\leq 4$~\cite{Experiment enhancement}. Therefore, even in weak magnetic
fields the cyclotron motion is preserved and the mixing of the LLs is small.
Note that the first term in $E_{\rm ex}$ linearly depends on the magnetic
field whereas $E_{\rm h}$ has an approximately quadratic dependence.

Let us now turn to the main subject of the paper, a partially filled upper
LL. Due to the electron-hole symmetry within one spin subband, it suffices to
consider only $0 < \bar{\nu}_N \leq \frac12$.

\begin{figure}
\centerline{
\psfig{file=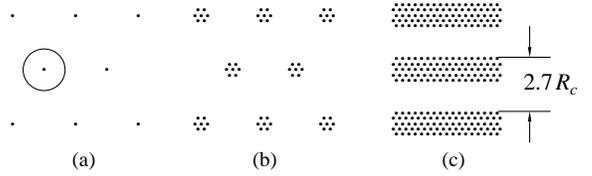,width=3in,bbllx=192pt,bblly=566pt,bburx=503pt,bbury=660pt}
}
\vspace{0.1in}
\setlength{\columnwidth}{3.2in}
\centerline{\caption{Guiding center density patterns.
(a) WC. One cyclotron orbit is shown. (b) ``Bubble'' pattern.
(c) Stripe pattern.
\label{Patterns}
}}
\end{figure}

We want to find the ground state of a partially filled LL. As we just saw,
the cyclotron motion is quantized. Thus, the remaining degrees of freedom are
associated with the guiding centers of the cyclotron orbits. In the ground
state these centers must arrange themselves in such a way that the
interaction energy is the lowest. This prompts a quasiclassical analogy
between the partially filled LL and a gas of ``rings'' with repulsive
interaction, the radius of each ring being equal to the cyclotron radius,
$R_c = \sqrt{2 N  + 1}\, l$. Strictly speaking, the guiding center can not be
localized precisely at one point, and so our analogy is not precise. However,
there exists a single-electron state~\cite{Kivelson} (so-called coherent
state) in which the guiding center has a very small scatter (of order $l$)
around a given point. (The wave-function for this state will be explicitly
given in Sec.~\ref{WC}). At large $N$ where $l \ll R_c$, the proposed analogy
becomes rather accurate.  Since the rings repel each other, it is natural to
guess that they form the WC. A trial wave-function for this state was written
by Aleiner and Glazman~\cite{Aleiner} by generalizing the Maki-Zotos $N
= 0$ wave-function~\cite{Maki} to arbitrary $N$:
\begin{equation}
 |\Psi\rangle = \prod_i c^\dagger_{\bbox{R}_i} | 0_N \rangle,
\label{Rings}
\end{equation}
where $|0_N\rangle$ stands for $N$ completely filled LLs,
$c^\dagger_{\bbox{R}}$ is the creation operator for a coherent
state~\cite{Kivelson}, and $\bbox{R}_i$ coincide with the lattice sites of
the classical WC with density $\bar{\nu}_N / (2\pi l^2)$.  When $\bar{\nu}_N$
is small, $\bar{\nu}_N \ll 1 / N$, the rings centered at neighboring lattice
sites do not overlap and the concept of the WC (Fig.~\ref{Patterns}a) is
perfectly justified. However, at larger $\bar{\nu}_N$ they overlap strongly.
In this work we show that at $\bar{\nu}_N \gg 1 / N$ the ground state is
completely different.  Generally speaking, the structure of the ground state
depends on $r_s$. In this Section we would like to announce the results only
for the practically important case, $r_s > 0.06$. We found that in the range
$1 / N \lesssim \bar{\nu}_N < \bar{\nu}_N^\ast$ where $\bar{\nu}_N^\ast$ is
somewhat smaller than $\frac12$, the electron form a ``super'' WC
(Fig.~\ref{Patterns}b) of large clusters (``bubbles'') containing about $\bar{\nu}_N N$
electrons each and separated by the distance that slowly changes from $2 R_c$
near the lower end of this range of $\bar\nu_N$ to approximately $3.3 R_c$
near the upper end. At larger $\bar{\nu}_N$, $\bar{\nu}_N^\ast < \bar{\nu}_N
\leq \frac12$, the clusters merge into parallel stripes
(Fig.~\ref{Patterns}c) with the spatial period approximately equal to $2.7
R_c$. The evolution of the ``bubble'' phase into the stripe one appears to be
continuous, and it is difficult to define $\bar\nu_N^\ast$ unambiguously. In
fact, $\bar\nu_N^\ast$ also depends on $N$.  At large $N$, however, it
approaches some limiting value, which we estimated to be about $0.4$.

Domain patterns shown in Fig.~\ref{Patterns}b and \ref{Patterns}c are well
known in many other physical and chemical systems~\cite{Seul}, examples of
the former being type I superconducting films in their intermediate state or
magnetic films. We will refer to the patterns in Fig.~\ref{Patterns}b
and~\ref{Patterns}c as the ``bubble'' and ``stripe'' phase, respectively.

It is, perhaps, surprising that the repulsive interaction leads to the
formation of compacted clusters. In Sec.~\ref{Quality} we study this
phenomenon in more detail and derive a general criterion for the interaction
potential to have this property.

%
%

The physical reason why the clusterization occurs in the considered system is
that the actual charge density variations in the ``super'' WC state are not
too large (of order $20\%$). This turns out to be possible because the CDW
period adjusts itself to the special ring-like shape of the electronic
wavefunctions.  In other words, the ground state is not a conventional charge
density but rather a guiding center density wave. Since the charge density
variations are not large, the increase in the electrostatic (Hartree) energy
due to the clusterization is also small. On the other hand, the
clusterization allows the system to achieve a lower value of the exchange
energy because the exchange interaction is rather short-range and each
particle has now more neighbors within this range.  Thus, it is not
accidental that the cohesive energy of our state
is of order of the exchange gap $E_{\rm ex}$. For example, the
cohesive energy at $\bar{\nu}_N \sim 1/2$ and $r_s \sim 1$ is given by
\begin{eqnarray}
E_{\rm coh}^{\rm CDW} \!&\approx&\displaystyle -\frac{r_s}{8 \sqrt{2} \pi}
\hbar\omega_c\ln\!\left(1 + \frac{0.3}{r_s}\right) -\frac{E_{\rm h}}{4}
\nonumber\\
\mbox{} &\approx& \displaystyle -0.01 \hbar\omega_c - \frac{E_{\rm h}}{4}.
\label{E_coh CDW}
\end{eqnarray}
The last line in this equation is obtained assuming that we are dealing with
the realistic case $r_s \sim 1$.

By the term cohesive energy we mean the difference in the energy per particle
at the upper LL in the ground state and in the UEL (appropriate at very high
temperature). The energy per particle in the UEL is given by $-\frac12
\bar{\nu}_N E_{\rm ex}$.

The CDW state turns out to be the most energetically favorable because it
represents the correlations on the largest length scale in the problem,
$R_c$. The correlations on the length scale $l$, built into the structure of,
say, the WC, are much less effective.  We believe that for the same reason at
large $N$ the Laughlin liquids can not compete with the CDW state either.

\begin{figure}
\centerline{
\psfig{file=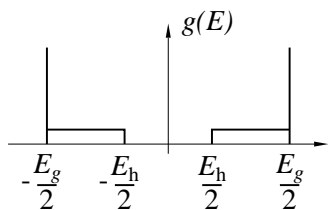,width=1.75in,bbllx=240pt,bblly=265pt,bburx=436pt,bbury=395pt}
}
\vspace{0.1in}
\setlength{\columnwidth}{3.2in}
\centerline{\caption{The DOS for the stripe CDW in the limit
of large $N$ (schematically).
\label{DOS}
}}
\end{figure}

We found that the most prominent feature of the one-particle density of
states (DOS) is two sharp peaks at the extremes of the spectrum (see
Fig.~\ref{DOS}).  Such peaks are the particular form of the van Hove
singularities.  The distance between the peaks for $\bar{\nu}_N \sim \frac12$
is equal to
\begin{equation}
E_{\rm g} \approx
\frac{r_s \hbar\omega_c}{\sqrt{2}\,\pi}\ln\!\left(1 + \frac{0.3}{r_s}\right)
+ E_{\rm h} \approx 0.07 \hbar\omega_c + E_{\rm h},
\label{Gap}
\end{equation}
which is very close to $E_{\rm ex}$.

Besides the peaks, the DOS has other structure, such as the shallow pseudogap
of width $E_{\rm h}$ centered at the Fermi energy. The existence of such a
gap was predicted in Refs.~\onlinecite{Aleiner,Aleiner_hydro}

Figure~\ref{DOS} depicts the asymptotical form of the DOS at truly large $N$.
At moderate $N$, the DOS appears merely as two distinct peaks.  The reason
for this is that the difference between $E_g$ and $E_{\rm h}$ is not too
large yet, while the van Hove singularities are not that extremely sharp. As
a result, the intervals $\frac12 E_{\rm h} < |E| < \frac12 E_g$ of constant
$g(E)$ (see Fig.~\ref{DOS}) are totally absorbed by the van Hove peaks.

%
%

The proposed ground state enables us to explain two interesting experimental
findings. One is a pseudogap in the tunneling DOS first observed in the
experiments with single quantum well~\cite{Ashoori} systems and, recently,
with double quantum well systems~\cite{Eisenstein,Turner}. It was found that
the differential conductivity as a function of the bias voltage exhibits two
maxima. The distance $E_{\rm tun}$ between the maxima appears to be linear in
magnetic field~\cite{Turner}.  We show that the dependence of $E_{\rm tun}$
on the magtetic field is more complicated. However, in the experimental range
$N \leq 4$, it can be satisfactorily fitted to the linear law $E_{\rm tun}
\approx 0.4\hbar\omega_c$, which compares favorably with the experimental
value of $0.45\hbar\omega_c$~\cite{Turner}.

Another important application of the proposed picture concerns with the
conductivity peak width of the integer quantum Hall effect in high-mobility
structures. It is usually assumed that the disorder in such systems is
long-range. In this case a semiclassical electrostatic model of
Efros~\cite{Efros} predicts that the electron liquid is compressible in a
large fraction of the sample area. If compressible liquid is considered to be
metallic, then the conductivity peaks are necessarily wide~\cite{Efros},
which is indeed the case at relatively high temperatures~\cite{Stormer}.
However, it is well-known that at low temperatures the peaks are narrow,
which may be interpreted as the pinning of the compressible liquid by the
disorder~\cite{Chklovskii}. The crystalline structure of the compressible
liquid (Fig.~\ref{Patterns}) makes such a pinning possible even though the
disorder is long-range. When the compressible liquid is pinned, it can not
move as a whole. As a result, at zero temperature the dissipative
conductivity $\sigma_{\rm xx}$ vanishes at all $\bar\nu_N \neq \frac12$.
Precisely at $\bar\nu_N = \frac12$, however, another mechanism of the
transport becomes operational. It is related to the motion of electrons along
the boundaries of the domains with $\nu_N = 1$ and $\nu_N = 0$, or, in other
words, along the ``bulk edge states''. The dc transport is possible only when
the bulk edge states percolate through the sample. This is realized only at
$\bar\nu_N = \frac12$, because the long-range order of the stripe phase is
destroyed by disorder. This explains narrow peaks of $\sigma_{\rm xx}$ at
zero temperature at half-integer $\nu$'s~\cite{Comment on peaks}.

At nonzero temperature the peaks have a finite width due to a hopping between
spatially close $\nu_N = 1$ domains. However, the consideration of such a
transport goes beyond the scope of the present paper.

The outline of the paper is as follows. In Sec.~\ref{Quality} we present the
qualitative discussion where we show that even perfectly repulsive
interaction may cause the clusterization of particles. In Sec.~\ref{CDW} we
formulate the self-consistent HF problem and give its approximate solution
under two kinds of simplifying assumptions (one corresponds to ``stripes''
and the other to ``bubbles''). In Sec.~\ref{num res} we report the results of
a numerical study of CDW patterns based on the trial
wave-function~(\ref{Rings}). In Sec.~\ref{Tunnel} we discuss the implications
of the CDW state for the double-well tunneling experiments. Finally,
Sec.~\ref{Conclusion} is devoted to conclusions.  Various details of
calculations, e.g., a careful comparison of the energies of the CDW and the
conventional WC are transferred to Appendix.
\section{Qualitative discussion}
\label{Quality}

Our results can be understood by analyzing the following toy model. Consider
a one-dimensional (1D) lattice gas interacting via the box-like potential
\begin{equation}
                          u(x) = u_0 \Theta(2 R - |x|)
\label{Box potential}
\end{equation}
and situated on the background of the same average density, interaction with
which is described by the potential of the same type but with the opposite
sign. One can say that each particle has a negative unit charge while the
background is charged positively.  We assume that $R$ is much larger than the
lattice constant $a$. We also assume that a multiple occupancy of the sites
is forbidden, then the average occupancy or the average filling factor
${\bar{\nu}}$ is always between $1$ and $0$.  Let us focus on the case
${\bar{\nu}} = \frac12$.

\begin{figure}
\centerline{
\psfig{file=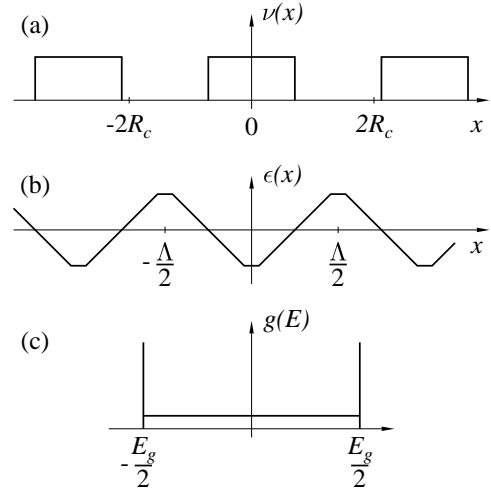,width=2.5in,bbllx=183pt,bblly=203pt,bburx=497pt,bbury=514pt}
}
\vspace{0.1in}
\setlength{\columnwidth}{3.2in}
\centerline{\caption{The CDW in the lattice gas.
(a) Local filling factor $\nu(x)$.
(b) One-particle energy $\epsilon(x)$.
(c) DOS in the lattice gas.
\label{Plot gas}
}}
\end{figure}

One of the possible particle distributions is the WC, i.e., the state where
every other lattice site is occupied. It can be shown that the absolute value
of its cohesive energy does not exceed $u_0$, the maximum value of the
two-particle interaction potential. Now we demonstrate that at $\bar{\nu} = 1
/ 2 $ the arrangement of the particles in a series of equidistant large
clusters of width $\sim R$ allows the system to attain the cohesive energy
as small as
\begin{equation}
                  E_{\rm coh} = -(3 - 2\sqrt{2}) (R/a) u_0.
\label{E_coh gas}
\end{equation}
For the obvious reason we call this state the CDW state. Since the spatial
period $\Lambda \sim R$ of this state is much larger than the average
interparticle distance, it is convenient to switch from the description in
terms of discreet particles to the continual representation where one uses
the local filling factor $\nu(x)$. We consider the CDW with the box-like
profile of $\nu(x)$:
\begin{equation}
         \nu(x) = \Theta\left(\frac{\Lambda}{4} - |x|\right), \quad
                    -\frac{\Lambda}{2} < x < \frac{\Lambda}{2},
\label{Box nu}
\end{equation}
see Fig.~\ref{Plot gas}a. The lowest value of the energy quoted above is
reached at $\Lambda = 2\sqrt{2}\, R \approx 2.8 R$.  Indeed, this period is
much larger than the average interparticle distance $2 a$, i.e., instead of
being equidistant as in the WC, the particles are compacted in large
clusters of the highest possible density $a^{-1}$.

The clusterization is advantageous because in contrast to, e.g., the usual
Coulomb law, here the interaction potential ceases to increase at distances
smaller than $2 R$. Therefore, particles can be brought closer to each other
at no energy cost. At the same time the particles in the interior of a given
cluster avoid the particles in the other clusters. Hence, they interact only
with the charge $\Lambda / 2 a$ of their own cluster.  Now recall that each
particle interacts with the background as well. The amount of the background
charge involved in this interaction is $(2 R / a) > (\Lambda / 2 a)$.
Therefore, the interaction with the positive background dominates and each
cluster resides in a deep potential well. The cohesive energy of the system
is determined simply by the average depth of this well.  Let us now calculate
the cohesive energy and then minimize it with respect to $\Lambda$. This way
we will find the optimal period of the state.

Define the one-particle energy $\epsilon(x)$ at a point $x$ by
\begin{equation}
  \epsilon(x) = \int\! \frac{{\rm d} x'}{a} u(x - x') [\nu(x') - \bar{\nu}],
\label{epsilon gas}
\end{equation}
then the cohesive energy is determined by the average value of $\epsilon(x)
\nu(x)$:
\begin{equation}
E_{\rm coh} = \frac{1}{2 \bar\nu}
\left\langle \epsilon(x) \nu(x) \rangle\right.
\label{E_coh vs epsilon}
\end{equation}
It is easy to see that  for $\nu(x)$ given by Eq.~(\ref{Box nu}) and
$\Lambda$ in the range $\frac83 R < \Lambda < 4 R$, $\epsilon(x)$ has an
approximately saw-tooth form (Fig.~\ref{Plot gas}b) and oscillates
between $\pm \frac12 E_g$, where
\begin{equation}
                      E_g = u_0 \frac{4 R - \Lambda}{a}.
\label{E_g gas}
\end{equation}
The cohesive energy can be readily evaluated to be $E_{\rm coh}(\Lambda) =
-E_g (\Lambda - 2 R) / 2 \Lambda$, which reaches its lowest value~(\ref{E_coh
gas}) at $\Lambda = 2 \sqrt{2}\,R$.

%
%

Let us now calculate the DOS $g(E)$, which we define by
\begin{equation}
   g(E) = \frac{1}{L_x} \int\! {\rm d}x \delta(\epsilon(x) - E),
\end{equation}
where $L_x$ is the length of the system. The integration yields
\begin{equation}
  g(E) = \frac{2}{\Lambda}\left\{
\begin{array}{cc}
\displaystyle\left|\frac{{\rm d}\epsilon}{{\rm d}x}\right|^{-1},&
|E| < \frac12 E_g\\
0, & |E| > \frac12 E_g,
\end{array}
\right.
\label{g}
\end{equation}
where the coordinate $x$ appearing on the upper line of Eq.~(\ref{g}) is any
of those where $\epsilon(x) = E$. Clearly, $g(E)$ is constant in the interval
$-\frac12 E_g < E < \frac12 E_g$ but diverges at the end points of this
interval (Fig.~\ref{Plot gas}d). These divergencies are the particular
form of the van Hove singularities inherent to the DOS of all periodic
structures.

Let us now return from the toy model to the real electrons at the upper
partially filled LL. Consider again the case $\bar\nu_N = \frac12$. As
discussed in Sec.~\ref{Introduction}, at such $\bar\nu_N$ the stripe CDW
pattern forms. In this case the problem is effectively 1D because the
one-electron basis states can be chosen in such a way that they are labelled
by one quantum number (the guiding center coordinate) $X$. In the Landau
gauge $\bbox{A} = -B x \hat{\bbox{y}}$, the wave-function of one of such
states is given by
\begin{equation}
\psi_X =
\frac{{\rm e}^{{\rm i}y X / l^2}}{\pi^{1/4} \sqrt{2^N N!\,l L_y}}
\exp\left[ -\frac{(x - X)^2}{2 l^2}\right] H_N\!\left(\frac{x - X}{l}\right),
\label{Stick}
\end{equation}
where $L_y$ is the $y$-dimension of the system and $H_N(x)$ is the Hermite
polynomial~\cite{Gradshtein}. The wave-function~(\ref{Stick}) is extended in
the $y$-direction but has a finite spread of $2 R_c$ in the $x$-direction.
Stricktly speaking, the HF potential $u_{\rm HF}(x)$ via which the basis
states interact, is different from the one given by Eq.~(\ref{Box
potential}). However, as it will be shown in the next Section, for $r_s \sim
1$ this potential is roughly equivalent to the following one
\begin{eqnarray}
\displaystyle u_{\rm HF}^{\rm eff}(x) &=& a \frac{\hbar\omega_c}{2 \pi^2 R_c}
\Theta(2 R_c - |x|) - a E_{\rm h} \delta(x),
\label{u_HF_eff}\\
a &=& \frac{2 \pi l^2}{L_y}.
\label{a}
\end{eqnarray}
Such a formula for $u_{\rm HF}^{\rm eff}(x)$ results from the particular form
of the bare interaction potential $v(r)$ (the interaction potential of two
point-like charges) in the 2DEG in a weak magnetic field, which is as
follows.  At very short $r \ll a_{\rm B}$ and very large $r \gg R^2_c /
a_{\rm B}$ distances, $v(r)$ coincides with the usual Coulomb law $v(r) = e^2
/ \kappa r$. At intermediate distances, it is significantly smaller than the
Coulomb potential because of a strong screening by the large number of
electrons at the lower LLs. Very crudely, $v(r)$ can be approximated by
\begin{equation}
 v(\bbox r) = \pi e^2 a_{\rm B} \delta(\bbox r) + E_{\rm h},
\quad r \lesssim 2 R_c,
\label{v r}
\end{equation}
where $E_{\rm h} = (e^2 a_{\rm B} / \kappa R_c^2)\ln (R_c / \sqrt{2}\, a_{\rm
B})$ [cf.~Eq.~(\ref{E_h})].

Clearly, the first term in $v(r)$ gives a non-vanishing contribution to the
interaction potential $u_{\rm HF}(x)$ between two basis states only if the
densities of the two states overlap, i.e., at $|x| < 2 R_c$. Beyond $2 R_c$,
this contribution becomes very small. This is represented by the first term
in Eq.~(\ref{u_HF_eff}), which exhibits a step-like discontinuity at $|x| = 2
R_c$.

The second term in $u_{\rm HF}^{\rm eff}(x)$ comes from the second
(``hydrodynamic'') term in the bare interaction potential. It is important
that this second term in $v(r)$ is almost constant in the real space up to
rather large distances (of the order of several $R_c$).
%
%
It is then clear on physical grounds that this long-range ``hydrodynamic''
term in $v(r)$ (or its image $a E_{\rm h}\delta(x)$ in $u_{\rm HF}^{\rm
eff}$) has no effect on the short length scale structure of the ground state.
In other words, as long as the characterisctic spatial scales of a given
state are of the order of $R_c$ or shorter, the contribution of the
``hydrodynamic'' term to the cohesive energy of such a state is the same. It
can be shown that this contribution is equal to
\begin{equation}
          E_{\rm coh}^{\rm h} = -\frac{1 - \bar\nu_N}{2} E_{\rm h}
\label{E_coh_h}
\end{equation}
(see also Sec.~\ref{CDW}). Therefore, the ground state structure is
determined by the box-like part and is exactly the same as in our toy model
(see Fig.~\ref{Plot gas}a).

\begin{figure}
\centerline{
\psfig{file=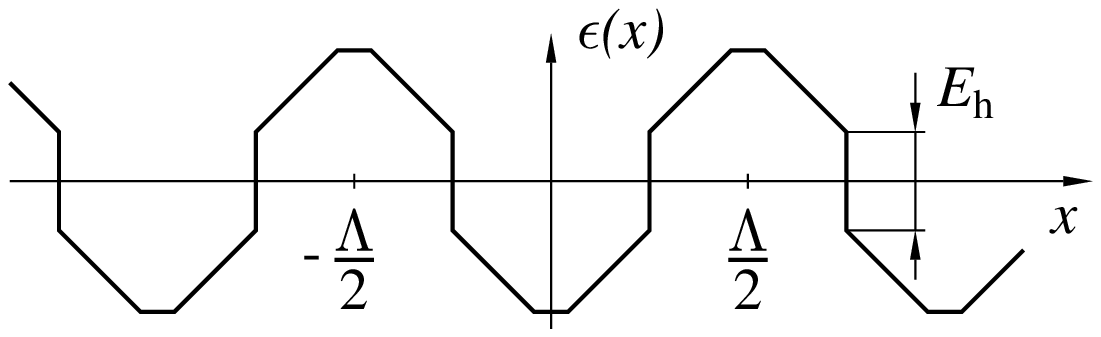,width=2.5in,bbllx=180pt,bblly=405pt,bburx=500pt,bbury=505pt}
}
\vspace{0.1in}
\setlength{\columnwidth}{3.2in}
\centerline{\caption{One-particle energy $\epsilon(x)$ for
the CDW in the real system (schematically).
\label{Plot real}
}}
\end{figure}

Given the results of the toy model, we can immediately derive the quantities
of interest for the real system (in the practically important case $r_s \sim
1$) as well.  First, the optimal CDW period $\Lambda$ should be close to $2.8
R_c$. Indeed, we found the value of $2.7 R_c$ for this quantity. Second,
Eq.~(\ref{E_coh CDW}) for the cohesive energy follows from Eq.~(\ref{E_coh
gas}) after the appropriate substitutions for $u_0$ and $R$ by the parameters
from Eq.~(\ref{u_HF_eff}). [One should not forget here to add the
contribution of the short-range part given by Eq.~(\ref{E_coh_h})].  Finally,
to deduce the functional form of the DOS, let us examine the effect of this
short-range part on $\epsilon(x)$. Clearly, it is to lower $\epsilon(x)$ by
$\frac12 E_{\rm h}$ in $\nu(x) = 1$ intervals and to raise it by the same
amount in the other intervals where $\nu(x) = 0$.  This generates the jumps
in $\epsilon(x)$ at $x = \pm \frac14\Lambda$ (Fig.~\ref{Plot real})
superimposed on the familiar saw-tooth profile of $\epsilon(x)$ in the
lattice gas model (Fig.~\ref{Plot gas}b).  Hence, the effect of the
short-range part on the DOS is to insert a hard gap of width $E_{\rm h}$
centered at zero energy (Fig.~\ref{DOS}).  Therefore, $E_g$ is augmented by
the same value, which is accounted for by the second term in Eq.~(\ref{Gap}).
The first term in this equation (for $r_s \sim 1$) follows from Eq.~(\ref{E_g
gas}) upon the appropriate substitutions for $u_0$ and $R$.

%
%

Note that in the real system, unlike in the toy model, the van Hove
singularities at the edges of the spectrum are not delta-function-like but
the inverse square-root ones. Indeed, the DOS is inversely proportional to
${\rm d}\epsilon / {\rm d}x$; hence, the type of the singularity at, say, $E
= -\frac12 E_g$ is determined by how this derivative goes to zero at $x \to
0$. The real interaction potential is never an ideally flat-top box, and in
reality the extrema in $\epsilon(x)$ in Fig.~\ref{DOS} are, in fact,
somewhat rounded. We expect that the second derivative ${\rm d}^2\epsilon /
{\rm d}x^2$ is finite at $x = 0$, which corresponds to the inverse
square-root singularity in $g(E)$.

Our toy model enabled us to derive our main results [Eqs.~(\ref{E_coh
CDW},\ref{Gap})] for the most practical case of $r_s \sim 1$. At
smaller $r_s$ the potential is no longer box-like and we have to examine the
problem more carefully, which will be done in the next Section.  Before we do
so, it is important to identify a general criterion, which enables one to tell
whether or not a given type of the repulsive interaction would lead to a
similar clusterization of particles. This criterion can be obtained by
analyzing the Fourier transform of the potential and is well-known in the
theory of the CDW systems.

For simplicity, let us return to the 1D case. The cohesive energy can be
written as
\begin{equation}
 E_{\rm coh} = \frac{1}{2 L_x \bar{\nu} a^2}
               \sum_{q \neq 0} \tilde{u}(q) \tilde{\nu}^2(q),
\end{equation}
where $L_x$ is the length of the system. (From now on, everywhere where using
the same symbol for both real space and $q$-space quantities may lead to
ambiguity, the Fourier transformed quantities are denoted by tildes.)
Obviously, if $\tilde{u}(q)$ is negative at certain $q$, then the formation
of a CDW with such $q$ will lead to the lowering of the system energy.
Therefore, the criterion for the CDW instability is the presence of {\em
negative values} of $\tilde{u}(q)$. The Fourier transform of the box-like
potential~(\ref{Box potential}) is given by $\tilde{u}(q) = 2\sin(2 q R) /
q$, which is indeed negative at certain $q$.  Note also that the CDW
instability is the strongest at $q = 3\pi / 4 R$ where $\tilde{u}(q)$ reaches
its lowest value. This particular $q$ corresponds to the spatial period of
$\frac83 R \approx 2.67 R$, which is very close to the optimal period of
approximately $2.8 R$ we found above. The reason why they are slightly
different is that the CDW instability enters the non-linear regime.
Apparently, a small change in the period enables the CDW to incorporate the
contribution of subleading harmonics in a more optimal way.

In the next Section we give details of the HF approximation from which the
true form of the interaction potential can be found.
\section{CDW state}
\label{CDW}

In this Section we formulate the self-consistent HF problem and give its
approximate solution.

The first-principle formulation of the HF problem involves into the
consideration all occupied $N + 1$ LLs together with several low-lying empty
ones. In other words, it deals with of order $N$ different species of
fermions interacting via of order $N^2$ different types of interaction
amongst themselves. Although for moderate $N$ such a treatment is
feasible, the solution can be obtained only numerically (see, e.g., MacDonald
and Aers~\cite{MacDonald}).

A different approach was put forward by Aleiner and Glazman~\cite{Aleiner}.
They showed that at sufficiently large $N$, $N \gg r_s^{-1} \gg 1$, the
degrees of freedom associated with lower $N$ LLs can be integrated out, and
derived an effective Hamiltonian~\cite{Aleiner} governing the low-energy
physics of the 2D liquid in a weak magnetic field:
\begin{equation}
\hat{H}_{\rm eff} = \frac{1}{2 L_x L_y}\sum_{\bbox{q}}
{\rho}(\bbox{q}) \tilde{v}(\bbox{q}) {\rho}(-\bbox{q}),
\label{H_eff}
\end{equation}
where ${\rho}(\bbox{q})$ is the projection of the density operator onto
the upper LL and
\begin{equation}
            \tilde{v}(q)=\frac{2\pi e^2}{\varepsilon(q) q}
\label{v}
\end{equation}
is the renormalized interaction potential (remind that the tilde is used for
Fourier transformed quantities). Physically, the bare Coulomb interaction
amongst the electrons at the upper partially filled LL gets renormalized
because electric fields become screened by the lower completely filled LLs.
The quantity $\varepsilon(q)$ has, therefore, the meaning of the dielectric
constant for the system of the filled LLs. It is given by~\cite{Aleiner}
%
%
\begin{equation}
\varepsilon(q) = \kappa
   \left\{1 + \frac{2}{q a_{\rm B}} \left[1 - J_0^2(q R_c)\right]\right\},
\label{epsilon}
\end{equation}
where $\kappa$ is the background dielectric constant and $J_0$ is the Bessel
function of the first kind~\cite{Gradshtein}. Note that the asymptotic
expressions for $\varepsilon(q)$,
\begin{equation}
\varepsilon(q) = \kappa\left\{
\begin{array}{ll}
\displaystyle 1 + \frac{2}{q a_{\rm B}},& R_c^{-1} \ll q \ll k_{\rm F}\\
\displaystyle 1 + \frac{R_c^2 q}{a_{\rm B}},& q \ll R_c^{-1},
\end{array}\right.
\end{equation}
were obtained earlier by Kukushkin {\it et al.\/}~\cite{Kukushkin}.
Eq.~(\ref{v r}) of the previous Section can be obtained by performing the
Fourier transform of Eq.~(\ref{v}).

Let us return to Eq.~(\ref{H_eff}). It contains the projected on the upper LL
density operator ${\rho}(\bbox{q})$ expressible in the form
\begin{equation}
{\rho}(\bbox{q}) = \sum_X F(q) {\rm e}^{-{\rm i} q_x X}
a^\dagger_{X_+} a \raisebox{-1.5pt}{\mbox{$_{X_-}$}},
\label{rho}
\end{equation}
where $a_{X}^\dagger$ ($a \raisebox{-1.5pt}{\mbox{$_X$}}$) is the creation
(annihilation) operators of the state~(\ref{Stick}), $X_\pm$ are defined by
$X_\pm = X \pm q_y l^2 / 2$, and $F(q)$, given by
\begin{equation}
F(q) = \int\! {\rm d} x {\rm d} y |\psi_X|^2 {\rm e}^{-{\rm i} q x},
\label{F def}
\end{equation}
bares the name of the form-factor of state~(\ref{Stick}). Performing the
integration, one obtains
\begin{equation}
F(q) = \exp \left( -\frac{q^2 l^2}4 \right)
               L_N\left( \frac{q^2 l^2}{2} \right),
\label{F}
\end{equation}
$L_N(x)$ being the Laguerre polynomial~\cite{Gradshtein}.
Following the usual procedure of the HF decoupling of the
Hamiltonian~(\ref{H_eff}) we get
%
%
\begin{equation}
\hat{H}_{\rm HF} = \frac{n_{\rm L}}{2}\sum_{\bbox{q}}
\tilde{u}_{\rm HF}(\bbox{q})
{\Delta}(-\bbox q)\sum_X {\rm e}^{-{\rm i}q_x X}
a^\dagger_{X_+} a \raisebox{-1.5pt}{\mbox{$_{X_-}$}},
\label{H_HF}
\end{equation}
where $n_{\rm L} = (2\pi l^2)^{-1}$ is the density of one completely filled
LL and
\begin{equation}
{\Delta} (\bbox q) = \frac{2 \pi l^2}{L_x L_y}
\sum_X {\rm e}^{-{\rm i}q_x X}
\langle a^\dagger_{X_+} a \raisebox{-1.5pt}{\mbox{$_{X_-}$}}\rangle
\label{Delta q}
\end{equation}
is the CDW order parameter~\cite{Fukuyama,Yoshioka}. By $\tilde{u}_{\rm HF}$
in Eq.~(\ref{H_HF}) we denote the HF potential, $\tilde{u}_{\rm HF}(q) =
\tilde{u}_{\rm H}(q) - \tilde{u}_{\rm ex}(q)$.  The Hartree potential
$\tilde{u}_{\rm H}(q)$ is given by
\begin{equation}
\tilde{u}_{\rm H}(q) = \tilde{v}(q) F^2(q).
\label{u_H}
\end{equation}
The exchange potential $\tilde{u}_{\rm ex}(q)$ in the reciprocal space turns
out to be proportional to the real-space Hartree potential,
\begin{equation}
\tilde{u}_{\rm ex}(q) = u_{\rm H}(q l^2) / n_{\rm L}.
\label{u_ex}
\end{equation}
From Eqs.~(\ref{F},\ref{u_H},\ref{u_ex}) and also from an asymptotic formula
for $F(q)$,
\begin{equation}
  F(q) \simeq J_0(q R_c),\quad  q \ll k_{\rm F},
\label{F asym}
\end{equation}
more convenient expressions for $\tilde{u}_{\rm H}(q)$ and $\tilde{u}_{\rm
ex}(q)$ at $R_c^{-1} \lesssim q \ll k_{\rm F}$ can be derived,
\begin{eqnarray}
& &n_{\rm L}\tilde{u}_{\rm H}(q) \approx
\frac{\hbar\omega_c}{2 + q a_{\rm B}}{J_0^2(q R_c)},
\label{u_H asym}\\
& &n_{\rm L}\tilde{u}_{\rm ex}(q) \approx
\frac{r_s \hbar\omega_c}{\sqrt{2}\,\pi}\nonumber\\
& & \mbox{} \times
\left\{\ln\!\left(\!1 + \frac {r_s^{-1}}{\sqrt{2}\,q R_c}\right) +
\frac{\sin(2 q R_c)}{2 q R_c [1 + (r_s / \sqrt{2})]} \right\} + E_{\rm h}.
\label{u_ex asym}
\end{eqnarray}

The cohesive energy can be obtained from Eq.~(\ref{H_HF}) by omitting the
wave-vector $q = 0$ in the sum, taking the quantum-mechanical average, and
then dividing the result by the total number of particles $\bar{\nu}_N n_{\rm
L} L_x L_y$ at the upper LL, which gives
\begin{equation}
E_{\rm coh} = \frac{n_{\rm L}}{2\bar{\nu}_N}
\sum_{\bbox{q} \neq 0}\tilde{u}_{\rm HF}(q)
\left|{\Delta}(\bbox{q})\right|^2.
\label{E_coh}
\end{equation}

We want to find the set of the CDW order parameters ${\Delta}(\bbox{q})$ that
minimizes the cohesive energy under certain conditions of the
self-consistency (see below) imposed on ${\Delta}(\bbox{q})$.  We will
present an approximate solution based on the consideration of only two
idealized CDW patterns: a system of uniform parallel stripes and a triangular
lattice of perfectly round ``bubbles''. As it was mentioned in the
Introduction, the stripe phase is more favorable in some interval of
$\bar\nu_N$ around the half-filling. Outside of this interval, it gets
replaced by the ``bubble'' phase. We will consider the two phases separately.

{\bf Stripe phase}. In this case ${\Delta}(\bbox{q})$ are non-zero only if
$q_y = 0$.  It is convenient to introduce the local filling factor ${\nu}_N$
(compare with $\bar{\nu}_N$, the {\em average\/} filling factor) by $\nu_N(x,
y) = \nu_N(x) \equiv \langle a^\dagger_x a
\raisebox{-0.5pt}{\mbox{$_x$}}\rangle$. In the stripe phase, the order
parameter $\Delta$ and $\nu_N$ are just proportional to each other:
\begin{equation}
                   \Delta = \frac{\nu_N}{L_x L_y}.
\label{Delta vs nu}
\end{equation}
The self-consistency condition mentioned above is
\begin{equation}
              \nu_N(x) = \Theta[\epsilon_{\rm F}-\epsilon(x)],
\label{Self-consistency}
\end{equation}
whose meaning is that all the states below the Fermi level $\epsilon_{\rm
F}$ are filled, and all others are empty. The self-consistent potential
$\epsilon(x)$ in Eq.~(\ref{Self-consistency}) is given by
\begin{equation}
\epsilon(x) = n_{\rm L}
\sum\limits_{q \neq 0} \tilde{u}_{\rm HF}(q)
{\Delta}(q \hat{\bbox{x}}) {\rm e}^{{\rm i} q x}.
\label{Self-energy}
\end{equation}
By $\hat{x}$ we mean the unit vector in the $x$ direction.
Equations.~(\ref{E_coh}-\ref{Self-energy}) define the HF
problem for the case of the unidirectional CDW.

For $N > 0$ the Hartree potential $\tilde{u}_{\rm H}(q)$ inevitably has
zeros due to the factor $F(q)$ containing the Laguerre polynomial
[Eq.~(\ref{u_H})].  The first zero, $q_0$, is approximately given by $q_0
\approx 2.4 / R_c$.  Since the exchange potential is always positive
[Eq.~(\ref{u_ex asym})], there exist $q$'s where the total HF potential
$\tilde{u}_{\rm HF}$ is negative. This leads to the CDW instability because
the energy can be reduced by creating a perturbation at any of such
wave-vectors (cf. Ref.~\onlinecite{Fukuyama}).
%
%
\begin{figure}
\centerline{
\psfig{file=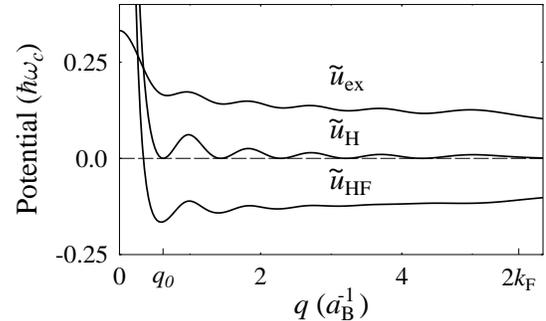,width=2.75in,bbllx=143pt,bblly=494pt,bburx=496pt,bbury=709pt}
}
\vspace{0.1in}
\nopagebreak
\setlength{\columnwidth}{3.2in}
\centerline{\caption{The Hartree, exchange, and HF potentials
in $q$-space for $N = 5$ and $r_s = 0.5$.
\label{Plot u_HF}
}}
\vspace{-0.1in}
\end{figure}

We will focus on the parameter range $0.06 < r_s < 1$ and $N < 50$, which
covers the cases of the experimental practice and even beyond. In such a
parameter range, the HF potential is negative at all wave-vectors $q > q_0$
and reaches its lowest value near $q = q_0$ (see Fig.~\ref{Plot u_HF}).  One
can guess then that the lowest energy CDW is the one with the largest
possible [under the conditions~(\ref{Self-consistency},\ref{Self-energy})]
value of $|{\Delta}(q_0 \hat{\bbox{x}})|$. The CDW having this property
consists of alternating stripes with filling factors $\nu_N(x) = 0$ and
$\nu_N(x) = 1$ (Fig.~\ref{Patterns}c). Within the class of unidirectional
CDWs we are considering now, this guess turns out to be correct. However, due
to the anharmonism of such a solution, the optimal spatial period $\Lambda$
of the CDW is slightly larger than $2 \pi / q_0$ and is equal to
\begin{equation}
                              \Lambda = 2.7 R_c.
\label{Lambda}
\end{equation}
Non-zero ${\Delta}(\bbox{q})$ for this solution are given by
\begin{equation}
 {\Delta}(q \hat{\bbox{x}}) = \frac{2}{\Lambda q}
\sin\left(\frac{\bar{\nu}_N}{2} \Lambda q\right)
\label{Delta box}
\end{equation}
provided $q$ is an integer multiple of $2\pi / \Lambda$.

Let us now derive Eq.~(\ref{Gap}) for the pseudogap. Clearly, $E_g = 2
|\epsilon(0)|$.  To calculate $\epsilon(0)$ we could, in principle, use
Eqs.~(\ref{u_H asym},\ref{u_ex asym}) to sum the series in
Eq.~(\ref{Self-energy}).  However, to establish the connection with the
previous Section, we will switch to the real space. Define a 1D HF potential
\begin{equation}
u_{\rm HF}(x) \equiv \frac{1}{L_y} \int\!{\rm d} q\, {\rm e}^{{\rm i}
q x}\, \tilde{u}_{\rm HF}(q),
\label{u_HF 1D}
\end{equation}
then $\epsilon(x)$ will be related to $\nu_N(x)$ in the way similar to
Eq.~(\ref{epsilon gas}):
\begin{equation}
  \epsilon(x) = \int\! \frac{{\rm d} x'}{a} u_{\rm HF}(x - x')
  [\nu_N(x') - \bar{\nu}_N],
\label{epsilon x}
\end{equation}
where $a$ is given by Eq.~(\ref{a}).

\begin{figure}
\vspace{0.1in}
\centerline{
\psfig{file=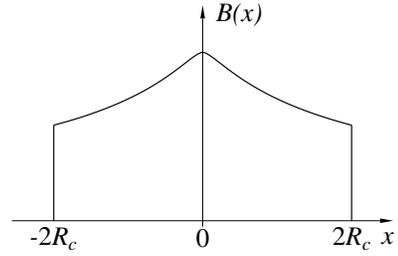,width=2in,bbllx=171pt,bblly=514pt,bburx=536pt,bbury=746pt}
}
\vspace{0.1in}
\setlength{\columnwidth}{3.2in}
\centerline{\caption{The function $B(x)$.
\label{Plot B(x)}
}}
\end{figure}

At large $N$ the potential $u_{\rm HF}(x)$ can be approximated by
\begin{eqnarray}
u_{\rm HF}^{\rm eff}(x) &=& a \displaystyle\frac{\hbar\omega_c}{2 \pi^2 R} B(x)
 - a E_{\rm h} \delta(x),
\label{u_HF 1D asym}\\
B(x) &=& \int\limits_0^{\pi / 2}\!
\displaystyle\frac{{\rm d}\varphi}{\left({r_s}/{\sqrt{2}}\right) +
\sqrt{1 - [1 - k^2(x)] \sin^2\varphi}},
\label{B}
\end{eqnarray}
where $k(x) \equiv x / 2 R_c$ (see Appendix~\ref{Derivation of B}).
Eq.~(\ref{u_HF 1D asym}) holds provided $x$ is larger than $a_{\rm B}$ but is
smaller than several $R_c$.  For $r_s \ll 1$ the function $B(x)$ satisfies
the following asymptotic relations:
\begin{equation}
B(x) = \left\{
\begin{array}{cc}
\displaystyle\ln\left(\frac{2 \sqrt{2}}{r_s}\right),& x = 0\\
\displaystyle K'\left(\frac{x}{2 R_c}\right),&
r_s R_c \lesssim |x| < 2 R_c\\
0, & |x| > 2 R_c,
\end{array}
\right.
\label{B asym}
\end{equation}
where $K'$ is the complete elliptic integral of the first
kind~\cite{Gradshtein}.  The plot of $B(x)$ is schematically shown in
Fig.~\ref{Plot B(x)}. One can see that it has a step-like discontinuity at $x
= 2 R_c$ already mentioned in Sec.~\ref{Quality}. In fact, at $r_s \sim 1$,
$B(x)$ is very nearly box-like, $B(x) \sim \Theta(2 R_c - |x|)$ and
Eq.~(\ref{u_HF_eff}) follows.

%
%

Using Eqs.~(\ref{epsilon x},\ref{u_HF 1D asym}), we find for $E_g$:
\begin{equation}
 E_g = \frac{\hbar\omega_c}{\pi^2 R_c}
\int\limits_0^{2 R_c}\!{\rm d}x\, B(x)\,
{\rm sgn}\left(x - \frac{\Lambda}{4}\right) + E_{\rm h},
\label{E_g}
\end{equation}
which with the help of Eq.~(\ref{B}) can be transformed into

\begin{equation}
 E_g \approx \hbar\omega_c \left(0.013 +
\frac{r_s}{\sqrt{2}\,\pi} \int\limits_{r_s}^{\Lambda / 8 R_c}
\frac{{\rm d} k}{k} \right) + E_{\rm h}.
\label{E_g II}
\end{equation}
After the substitution $\Lambda = 2.7 R_c$, one recovers Eq.~(\ref{Gap}). (The
small number in the parentheses of Eq.~(\ref{E_g II})
is the result of a numerical evaluation of a certain integral.
To obtain Eq.~(\ref{Gap}) it was disregarded).

Let us now calculate the cohesive energy [Eq.~(\ref{E_coh CDW})]. When $r_s$
is not much smaller than unity, $B(x)$ remains approximately box-like, and,
consequently, $\epsilon(x)$ in the $\nu_N(x) = 1$ intervals has an
essentially triangular cusp (Fig.~\ref{Plot real}).  Using Eq.~(\ref{E_coh
vs epsilon}), we then arrive at
\begin{equation}
                E_{\rm coh}^{\rm CDW} = -\frac18 (E_g + E_{\rm h}),
\end{equation}
which leads to Eq.~(\ref{E_coh CDW}).

Note that at smaller $r_s$, $r_s \sim 0.1$, the cusp in $\epsilon(x)$
deviates from the triangular form. This is illustrated in Fig.~\ref{Plot
DOS}, where $\epsilon(x)$ for $r_s = 0.1$, $N = 30$ is shown. In this
figure one can see the bending of the initially straight lines of
Fig.~\ref{Plot real}.

%
%
\begin{figure}
\vspace{0.1in}
\centerline{
\psfig{file=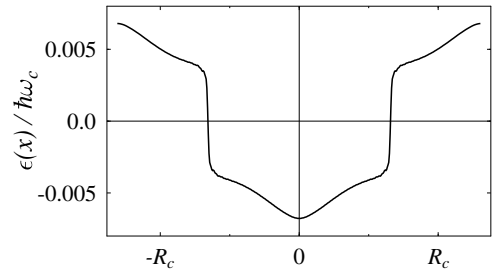,width=2.5in,bbllx=142pt,bblly=514pt,bburx=496pt,bbury=709pt}
}
\vspace{0.1in}
\setlength{\columnwidth}{3.2in}
\centerline{\caption{$\epsilon(x)$ for $r_s = 0.1$, $N = 30$.
\label{Plot DOS}
}}
\end{figure}

\begin{figure}
\vspace{0.1in}
\centerline{
\psfig{file=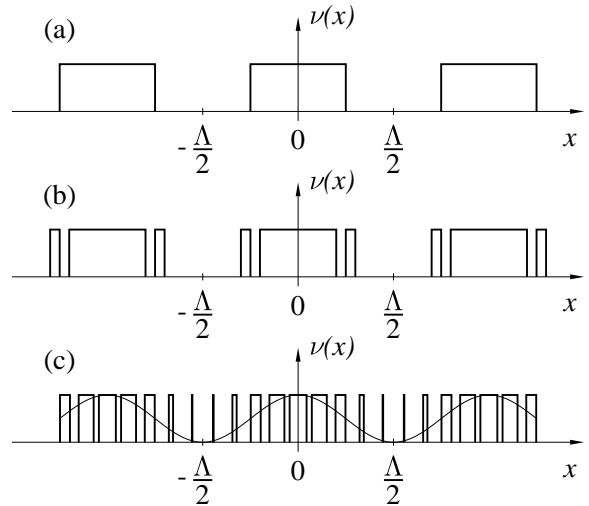,width=3in,bbllx=154pt,bblly=218pt,bburx=496pt,bbury=518pt}
}
\vspace{0.1in}
\setlength{\columnwidth}{3.2in}
\centerline{\caption{The local filling factor in the
stripe CDW pattern for different $r_s$. (a) $r_s > r_s^\ast$
(b) $r_s \sim r_s^\ast$ (c) $r_s \ll r_s^\ast$.
\label{Boxes}
}}
\vspace{-0.1in}
\end{figure}

This bending is due to the deviation of $B(x)$ from the ideally flat-top box,
or in physical terms, due to the Hartree term in the interaction. The Hartree
interaction reduces the slope of $\epsilon(x)$, or the magnitude of the
directed inward the stripes electric field $({{\rm d}\epsilon}/{{\rm d}x})
\hat{\bbox{x}}$. The magnitude of the field is equal to
\begin{equation}
\frac{{\rm d}\epsilon}{{\rm d}x} =
\frac{\hbar\omega_c}{2 \pi^2 R_c} \left[2 B\left(\frac{\Lambda}{2}\right) -
B(0) \right].
\label{Field}
\end{equation}
at the stripe boundary $x = \Lambda / 4$ (or, more precisely, at a distance
$\sim a_{\rm B}$ from the boundary). Eq.~(\ref{Field}) enables one to find
the range of $r_s$ where the simple CDW profile with one stripe per period
remains stable.  Indeed, there is a critical value of $r_s$ at
which the electric field near the stripe boundary vanishes.  Using
Eqs.~(\ref{Lambda},\ref{B asym},\ref{Field}), one arrives at
\begin{equation}
r_s^\ast = 2 \sqrt{2} \exp\left[- 2 K'\left(\frac{\Lambda}{4 R_c}\right)\right]
\approx 0.06
\label{r_s}
\end{equation}
for the critical $r_s$. At $r_s = r_s^\ast$, the original stripe breaks into
three smaller ones: the central one, almost as wide as the original stripe,
and two narrow ones (of width $\sim a_{\rm B}$) on the sides
(Fig.~\ref{Boxes}b).  Note that this transformation resembles the edge
reconstruction of quantum dots~\cite{Chamon} (the decrease of $r_s$ in this
analogy is equivalent to the steepening of the quantum dot confining
potential).

Since $r_s = 0.06$ is rather difficult to reach under the terrestrial
conditions, only the box-like solution and Eq.~(\ref{E_coh CDW}) have a
practical value. The next four paragraphs are devoted to a purely theoretical
issue of the CDW structure at truly small $r_s$. Unmotivated reader can skip
these paragraphs.

We expect that the reduction of $r_s$ beyond $r_s^\ast$ leads to further
increase in the number of the stripes per period. It is important also that
these stripes will be of unequal width, such that after a coarse-grain
averaging of the filling factor $\nu_N(x)$, it would appear approximately
sinusoidal (Fig.~\ref{Boxes}c). To understand that let us go back to the
$q$-space.

In the limit $r_s \ll r_s^\ast$ and $N \gg r_s^{-2}$, the exchange potential
$\tilde{u}_{\rm ex}(q)$ is on average much smaller than the Hartree potential
$\tilde{u}_{\rm H}(q)$ for $R_c^{-1} < q < (R_c r_s)^{-1}$ [see
Eqs.~(\ref{u_H},\ref{u_ex asym})], so the total HF potential is negative in
small $q$ intervals centered at the zeros of $\tilde{u}_{\rm H}(q)$. The
absolute minimum of $\tilde{u}_{\rm HF}(q)$ is still situated near $q = q_0$.
If one chooses the wave-vector of the principle harmonic to be $q_0$, already
the next harmonic of the box-like profile $q = 3 q_0$ will correspond to a
large positive $\tilde{u}_{\rm HF}(q)$.  Hence, to minimize the energy of the
system all such unfavorable harmonics must be suppressed.  In other words,
the CDW may be only slightly anharmonic. In a simplified description, the
spatial distribution of the filling factor is sinusoidal, and has the
amplitude $\bar\nu_N$ and the spatial period $2 \pi / q_0$.  Let us find the
cohesive energy of the stripe phase at given $\bar\nu_N$.  Since the filling
factor is approximately sinusoidal in $x$, in formula~(\ref{E_coh}) we have
to retain only two terms with $q_x = \pm q_0$ for which
$\left|{\Delta}(\bbox{q}) \right| \simeq \frac12\bar\nu_N$. Taking advantage
of Eq.~(\ref{u_ex asym}), we arrive at the estimate
\begin{equation}
E_{\rm coh}^{\rm CDW} \simeq
-\frac{\bar{\nu}_N r_s\hbar\omega_c}{4 \sqrt{2}\,\pi}
\ln\!\left(1 + \frac{0.3}{r_s}\right) -
\frac{1 - \bar{\nu}_N}{2} E_{\rm h}.
\label{E_coh CDW III}
\end{equation}
The last term was obtained by a more accurate procedure based on the sum
rule~\cite{Yoshioka}
\begin{equation}
  \sum_{\bbox{q} \neq 0} \left|{\Delta}(\bbox{q}) \right|^2 =
  \bar{\nu}_N (1 - \bar{\nu}_N),
\label{Sum rule}
\end{equation}
which is just another way to derive Eq.~(\ref{E_coh_h}).

%
%

In fact, the function $\nu_N(x)$ is more complicated. First, there are
deviations from the pure sinusoidal form in narrow regions centered at the
extrema of $\nu_N(x)$. In these regions $\nu_N(x)$ is flattened, such that it
reaches zero (or one) not at single points but in small intervals of $x$.
Second, even with these corrections, the simple sinusoidal of $\nu_N(x)$ is
not the complete answer yet because, taken literally, it contradicts to
Eq.~(\ref{Self-consistency}).  Indeed, according to
Eq.~(\ref{Self-consistency}), the local filling factor can be only one or
zero whereas we just argued that $\nu_N(x)$ takes intermediate values as
well. The contradiction is resolved by the fine structure of the CDW.
Namely, the CDW profile consists of many narrow boxes (Fig.~\ref{Boxes}c),
and appears sinusoidal only after the coarse-grain averaging. To find the
characteristic width of such boxes we have to analyze the HF potential more
carefully. It is easy to see that at wave-vectors $q > 1 / R_c r_s$, the
exchange potential becomes larger than the average of the oscillating Hartree
potential [Eqs.~(\ref{u_H},\ref{u_ex asym})]. Therefore, in this range of $q$
there are many harmonics, which need not to be suppressed. Thus, the typical
distance between the boxes is of the order of $\xi \equiv \sqrt{2}\, r_s
R_c$. The appearance of the scale $\xi$ in the ground state structure is not
accidental. It is related to the fact that $\xi$ is nothing else than the
range of the exchange potential $u_{\rm ex}(r)$ in the real space [at $r >
\xi$ $u_{\rm ex}(r)$ rapidly decreases:  $u_{\rm ex}(r) \propto 1 / r^2$].

{\bf ``Bubble'' phase}. We will consider only the large $N$ limit where $R_c
\gg l$ and, therefore, both the period of the bubble lattice and the radius
of the ``bubbles'' are much larger than the magnetic length. In this case we
can still use the concept of the local filling factor, but now it depends on
both $x$ and $y$ coordinates. The filling factor $\nu_N(x, y)$ is assumed to
be unity inside the ``bubbles'' and zero everywhere else. The ``bubbles''
form a triangular lattice. The relation between the lattice constant
$\Lambda_b$ and the radius $r_b$ of the ``bubbles'' is
\begin{equation}
\frac{r_b}{\Lambda_b} = \left(\frac{\sqrt{3}\,\bar\nu_N}{2 \pi}\right)^{\frac12}.
\label{r_b}
\end{equation}

Similar to the case of stripes, we will use an asymptotic expression for the
HF potential
\begin{equation}
\displaystyle n_{\rm L} u_{\rm HF}(\bbox{r}) \approx
\frac{\hbar\omega_c}{2 \pi^3 R_c}
\frac{\Theta(2 R_c - r)}{\sqrt{4 R_c^2 - r^2}} + E_{\rm h}
[n_{\rm L} - \delta(\bbox{r})],
\label{u_HF asym r}
\end{equation}
valid at $r_s \sim 1$ and $r$ smaller than several $R_c$. One possible way to
obtain this equation is to start from the formula for $u^{\rm eff}_{\rm HF}$
in $q$-space (see Sec.~\ref{Quality}) and then perform its 2D Fourier
transform.

\begin{figure}
\vspace{0.1in}
\centerline{
\psfig{file=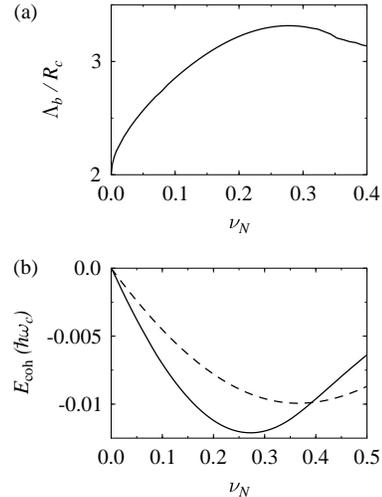,width=2in,bbllx=115pt,bblly=235pt,bburx=370pt,bbury=570pt}
}
\vspace{0.1in}
\setlength{\columnwidth}{3.2in}
\centerline{\caption{(a) The optimal distance $\Lambda_b$ between neighboring
bubbles as a function of the filling factor. (b) The cohesive energy of the
``bubble'' and stripe phases (solid and dashed lines, respectively).  The
plotted dependencies correspond to the limit $N \to \infty$, $r_s \sim 1$.
The ``hydrodynamic'' term in the cohesive energy, small in this limit, is
neglected.
\label{Bubbles}
}}
\end{figure}

With the help of Eqs.~(\ref{r_b},\ref{u_HF asym r}), for every given
$\bar\nu_N$ we have numerically calculated the cohesive energy of the
``bubble'' phase as a function of $\Lambda_b$. This way we have been able to
determine the optimal value of $\Lambda_b$, which is plotted in
Fig.~\ref{Bubbles}a. Note that at $\nu_N \lesssim 0.1$ the optimal period
closely follows the formula
\begin{equation}
\Lambda_b \simeq \frac{2 R_c}{1 - 2 (r_b / \Lambda_b)},
\label{Lambda_b}
\end{equation}
whose meaning is that the neighboring ``bubbles'' barely interact with
each other.

At $\bar\nu_N = \nu_N^\ast \approx 0.39$, the cohesive energies of the stripe
and the ``bubble'' phases become equal (Fig.~\ref{Bubbles}b). At smaller
$\bar\nu_N$, the ``bubble'' phase replaces the ``stripes''. Arguments can be
given that this transition is of the first order~\cite{Comment on order}.

The dominance of the ``bubble'' phase over the ``stripes'' at small
$\bar\nu_N$ allows a simple geometrical interpretation. Recall that at small
$\bar\nu_N$ the ``bubbles'' barely interact with each other. The situation in
the stripe phase is similar: the optimal period is very close to $\Lambda
\simeq 2 R_c / (1 - \bar\nu_N)$, so that only the particles within the same
stripe interact with each other. Given this, the cohesive energy is
determined by the interactions of particles within a single stripe (for the
stripe phase) or within a single ``bubble'' (for the ``bubble'' phase).
In the stripe phase, each particle interacts with all the particles within
the area $4 R_c \times \Lambda\bar\nu_N \simeq 8 R_c^2 \bar\nu_N$. In the
``bubble'' phase, the corresponding area is $\pi r_b^2 \simeq 2 \sqrt{3}\,
R_c^2 \bar\nu_N$, i.e., roughly by a factor of two smaller.  Thus, in the
``bubble'' phase the particles avoid each other more effectively, and this
phase should be more energetically favorable.  It is possible to further
elaborate along this way of reasoning and to show that for the interaction
potential~(\ref{u_HF asym r}), the ratio of the cohesive energies of the two
phases tends to $1.7$ as $\bar\nu_N \to 0$ in agreement with data of
Fig.~\ref{Bubbles}b.

%
%

On the intuitive level, the ``bubble'' phase is expected to appear as $\nu_N$
decreases because of the imminent Wigner crystallization at sufficiently
small $\bar\nu_N$. Indeed, the WC is a particular case of the ``bubble''
phase with $n_e = 1$, where $n_e$ is the number of particles in one
``bubble''. When $\bar\nu_N \gg 1 / N$, $n_e$ is large and can be found from
\begin{equation}
n_e = \frac{\sqrt{3}}{2 \pi} \left(\frac{\Lambda_b}{R_c}\right)^2 N \bar\nu_N.
\label{n_e}
\end{equation}
following from Eq.~(\ref{r_b}). As $\bar\nu_N$ becomes smaller, $n_e$
decreases. Eventually, at $\bar\nu_N \sim 1 / N$, $n_e$ reaches the value of
one and the ``bubble'' state becomes the ordinary WC (Fig.~\ref{Patterns}a).
Thus, the ``bubble'' phase appears as a natural intermediate state between
the stripe phase and the WC.

Equation~(\ref{n_e}) implies that in the interval $1 / N < \bar{\nu}_N \leq
\frac12$ the optimal number of electrons in one ``bubble'' is larger than
one, i.e., that the ``bubble'' phase is more energetically favorable than the
WC. This issue is discussed in more detail in Appendix~\ref{WC}.

So far, we have been discussing the case $r_s \sim 1$. We expect that with
decreasing $r_s$, the original ``bubbles'' break into smaller ones, similar to
the case of the stripes.  (Fig.~\ref{Boxes}b and~\ref{Boxes}c). The
characteristic distance between neighboring bubbles is of the order of $\xi
\sim \Lambda r_s$.  Such smaller ``bubbles'' contain fewer electrons each;
therefore, the transition to the WC phase occurs at larger filling factor
$\bar\nu_N \sim 1 / (N r_s^2)$.

The next Section is devoted to numerical simulations, which confirm the
formation of the  stripe and ``bubble'' phases.
\section{Numerical study}
\label{num res}

In the preceding Section we used the analytical arguments to show that the
electron liquid at the upper partially filled LL is unstable against the CDW
formation. At moderate (``realistic'') $r_s$, the CDW instability is very
strong and the resulting CDW content is highly anharmonic. In this situation
it is rather difficult to predict the optimal CDW pattern analytically.
However, in analogy to other physical systems~\cite{Seul}, one can anticipate
the stripe and ``bubble'' pattern (Fig.~\ref{Patterns}b and
\ref{Patterns}c) formation. Below we describe our finite-size numerical
simulations, which indeed demonstrate this.

The trial wave-function used in our modeling is given by Eq.~(\ref{Rings})
where the centers $\bbox{R}_i$ are chosen from the sites of the triangular
lattice. The lattice constant is equal to $(4\pi / \sqrt{3})^{1/2} l$ in
physical units. Hence, the fully populated lattice corresponds to the average
density $1 / (2 \pi l^2)$, i.e., to the filling factor $\bar\nu_N = 1$.
Denote by $n_i \in \{0, 1\}$ the occupancy of $i$-th site.  Let us derive the
expression for the cohesive energy of the trial state for a given set of
$n_i$'s.

First we examine the state with a single occupied site at the origin.  This
state bares the name of the coherent state~\cite{Kivelson}. We will remind
several properties of such a state.  The probability density distribution,
i.e., the square of the absolute value of the wave-function $\phi(\bbox{r})$
of the coherent state is given by
\begin{equation}
|\phi(\bbox{r})|^2 = \frac{1}{2 \pi l^2\, N!}
\left(\frac{r^2}{2 l^2}\right)^N {\rm e}^{-r^2 / 2 l^2}.
\label{Coherent state}
\end{equation}
It has a sharp maximum at $r = \sqrt{2 N + 1}\,l = R_c$, i.e., at the
location of the classical cyclotron orbit. The characteristic width of the
maximum in the radial direction is $l$. Thus, the picture of the electron
localized within a narrow ring naturally appears (see
Sec.~\ref{Introduction}).

Using Eq.~(\ref{Coherent state}), the order parameter [Eq.~(\ref{Delta q})]
of a single coherent state can be easily calculated. It is equal to
\begin{equation}
\Delta(q) = \Delta_c(q) \equiv \frac{1}{L_x L_y}
\exp\left(-\frac14 q^2 l^2\right).
\label{Delta coherent}
\end{equation}

Consider now two coherent states with centers at points $\bbox{R}_1$ and
$\bbox{R}_1$, which are separated by a distance $r = |\bbox{R_2} -
\bbox{R_1}|$. If $r \gg l$, such two states have a very high degree of
orthogonality [owing to oscillating phase factors not shown in
Eq.~(\ref{Coherent state})]. The overlap $A(r)$ between them is given by
\begin{equation}
A(r) \equiv \left|\left\langle
c \raisebox{-1.5pt}{\mbox{$_{\bbox{R}_1}$}} c_{\bbox{R}_2}^\dagger
\right\rangle\right|^2 = \exp\left(-\frac{r^2}{2 l^2}\right).
\label{A}
\end{equation}
As a result, with high accuracy, the order parameter of the HF
state~(\ref{Rings}) of two electrons is simply additive: $\Delta(\bbox{r})
\simeq \Delta_c(\bbox{r} - \bbox{R}_1) + \Delta_c(\bbox{r} - \bbox{R}_2)$.
This holds for many-electron state as well, provided that the guiding center
separation in each pair of electrons exceeds $l$.  Using
Eqs.~(\ref{E_coh},\ref{Delta coherent}), we arrive at
\begin{eqnarray}
E_{\rm coh} &\simeq & \frac{1}{2 N_e} \sum_{i\neq j} [(n_i - \bar{\nu}_N)
g_{\rm HF}( \bbox{R}_i- \bbox{R}_j) \nonumber\\
\mbox{} &\times & (n_j - \bar{\nu}_N)] - \frac{\bar\nu_N}{2} E_{\rm ex},
\label{E_coh num}
\end{eqnarray}
where $N_e$ is the total number of electrons, and the quantity $g_{\rm
HF}(r)$, defined through its Fourier transform,
\begin{equation}
  \tilde{g}_{\rm HF}(q) = \tilde{u}_{\rm HF}(q) {\rm e}^{-\frac14 q^2 l^2},
\label{g_HF}
\end{equation}
has the transparent meaning of the interaction energy of two coherent states
whose centers are separated by the distance $r$. In the actual simulations,
we replaced $g_{\rm HF}(r)$ by
\begin{equation}
               G_{\rm HF}(r) = \frac{g_{\rm HF}(r)}{1 - A(r)}
\label{G_HF}
\end{equation}
to take into account a nonzero overlap $A(r)$.  In fact, the overlap is not
too small only for nearest lattice sites, for which $A \approx
0.027$~\cite{Comment on many-body}.

\begin{figure}
\vspace{0.1in}
\centerline{
\psfig{file=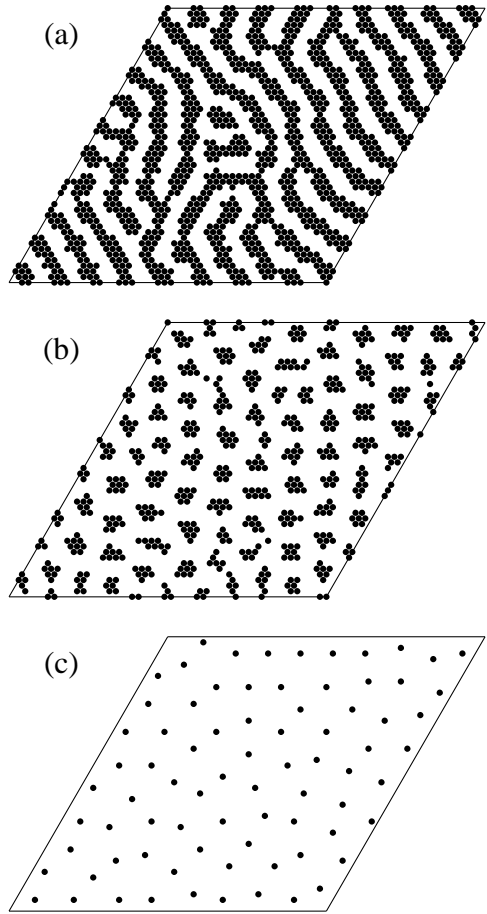,width=2.5in,bbllx=182pt,bblly=142pt,bburx=441pt,bbury=632pt}
}
\nopagebreak
\vspace{0.1in}
\nopagebreak
\setlength{\columnwidth}{3.2in}
\centerline{\caption{The CDW patterns produced by our numerical simulations
with the parameters $r_s = \protect\sqrt{2} / 3$, $N = 10$ for three
different values of $\bar\nu_N$:  (a) $\bar\nu_N = \frac12$
(b) $\bar\nu_N = \frac14$ (c) $\bar\nu_N = \frac{1}{16}$.
\label{Numericals}
}}
\end{figure}

In the simulations the lattice had the form of a parallelogram (see
Fig.~\ref{Numericals}) and contained altogether $50 \times 50$ sites.  In
every run, the goal of the simulations was to minimize the energy~(\ref{E_coh
num}) with respect to different configurations of the given number of the
occupied sites, i.e., at given $\bar\nu_N$. Since the total number of such
configurations is enormous even for a relatively small lattice, the true
ground state is extremely hard to find.  Fortunately, the
Hamiltonian~(\ref{E_coh num}) is exactly of the form studied in the context
of electrons localized at charged impurities in doped
semiconductors~\cite{Efros_Shklovskii}, and we can employ some techniques of
finding the approximate solution, developed in that field.

Our computational procedure starts from some initial configuration of the
occupied sites (usually a random one). At each step the configuration is
changed in favor of a new one with a lower energy. Ideally, the algorithm has
to contain many stages with different rules to pick up the new configuration
at each stage.  At the first stage the new configuration differs from the
previous one by the position of only one occupied site; at the second stage,
by the positions of the two, and so on.  We, however, restricted ourselves to
the first stage only. Presumably, it already gives a good approximation to
the ground state. The site to move is chosen according to the following
procedure. First, we calculate the potentials $\epsilon_i$ of all
the lattice sites,
\begin{equation}
\epsilon_i = N_e \frac{\partial E_{\rm coh}}{\partial n_i} = \sum_{j\neq i}
G_{\rm HF}(\bbox{R}_i - \bbox{R}_j) (n_j -\bar{\nu}_N),
\label{epsilon_i}
\end{equation}
and find the occupied site $i$ with the highest potential. Then we scan all
the vacant sites $j$, calculating the quantity
\begin{equation}
\delta E_{i\rightarrow j} = \epsilon_j - \epsilon_i -
G_{\rm HF}(\bbox{R}_i - \bbox{R}_j),
\label{change_of_energy}
\end{equation}
which is the change in the system energy upon the relocation of the occupied
site $i$ to the vacant site $j$~\cite{Efros_Shklovskii}. The relocation is
performed to the vacant site with the largest positive $\delta
E_{i\rightarrow j}$. If all $\delta E_{i\rightarrow j}$ for the given $i$ are
negative, then we try to relocate another occupied site $i$ in the same
manner. Eventually, if the pair of an occupied site $i$ and a vacant site $j$
with positive $\delta E_{i\rightarrow j}$ can not be found, the algorithm
terminates.

The results of the calculations with the parameters $r_s = \sqrt{2}/3$,
$N=10$, $\bar{\nu}_N = \frac12, \frac14, \frac{1}{16}$ are shown in
Fig.~\ref{Numericals}. In Fig.~\ref{Numericals}a one can see that at
$\bar{\nu}_N = \frac12$ the stripe pattern forms.  The deviations from the
ideal picture of identical parallel stripes are mainly due to the
incommensurability of the lattice constant with the optimal CDW period.
Other factors working in the same direction are the finite size of the
lattice and the fact that the algorithm is able to find only an approximation
to the ground state. Due to the same reasons, it is difficult to pinpoint the
transition to the ``bubble'' phase. However, we can put some bounds on it.
For example, at $N = 10$, the transition occurs within the interval $0.3 <
\bar{\nu}_N < 0.4$ (in agreement with our earlier estimate $\nu_N^\ast =
0.39$). At smaller $\bar\nu_N$, the pattern of isolated ``bubbles'' becomes
fully developed, see Fig.~\ref{Numericals}b.

We found that both the distance between the ``bubbles'' ($\Lambda_b \approx
3.3 R_c$ at $\bar{\nu}_N = \frac14$) and the average number of electrons in
one ``bubble'' $n_e \approx 3 \bar{\nu}_N N$ are in agreement with the
asymptotical laws given by Eq.~(\ref{n_e}) and the data of
Fig.~\ref{Bubbles}. As $\bar\nu_N$ goes down, $n_e$ becomes smaller, and, at
sufficiently small filling factor ($\bar\nu_N \sim 0.1$ for $N = 10$), the
``bubbles'' consist of only single occupied sites. At this moment the
distinction between the CDW and the WC disappears. At even smaller
$\bar{\nu}_N$, the occupied sites become more distant
(Fig.~\ref{Numericals}c).

Summarizing the results discussed in this Section, we see that our numerical
simulations give an additional piece of evidence in favor of the proposed CDW
ground state. In the next Section we discuss the one-particle DOS of the CDW
state and its relation to the recent tunneling experiments~\cite{Turner}.

%
%
%
\section{Tunneling properties}
\label{Tunnel}

In Sec.~\ref{Quality} we found that the one-particle DOS consists, roughly, of
the two van Hove singularities at the extremes of the spectrum $E = \pm
\frac12 E_g$ with $E_g$ given by Eq.~(\ref{Gap}).  Experimentally, the DOS
can be probed by the double-well tunneling
experiments~\cite{Eisenstein,Turner}. We derive the expression for the
tunneling conductance as a function of the voltage difference between the two
wells and then compare this expression with the experimental results of
Ref.~\onlinecite{Turner}.

\begin{figure}
\vspace{0.1in}
\centerline{
\psfig{file=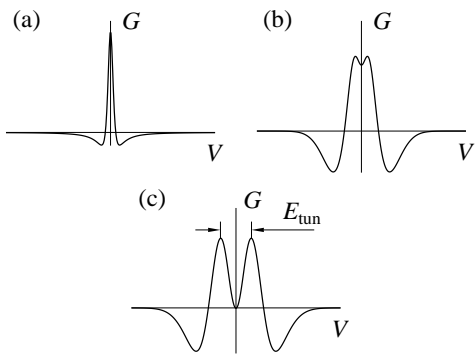,width=2.5in,bbllx=165pt,bblly=305pt,bburx=490pt,bbury=550pt}
}
\vspace{0.1in}
\setlength{\columnwidth}{3.2in}
\centerline{\caption{The evolution of the peak in the differential
conductivity $G(V)$ as the magnetic field increases.
(a) Shubnikov-de Haas regime.
(b) Spin-unresolved QHE regime.
(c) Spin-resolved QHE regime.
\label{Peaks}
}}
\end{figure}

In the tunneling experiments of Ref.~\onlinecite{Turner,Eisenstein} two about
$200{\rm\AA}$ thick high-mobility GaAs quantum wells, each containing the
2DEG, are separated by an $\rm AlGaAs$ barrier of width about $150{\rm\AA}$.  The
experiment consists in measuring the low-temperature current-voltage
characteristics of the double-well system in the magnetic field applied
perpendicular to the 2D planes. The results of the experiments (illustrated
by Fig.~\ref{Peaks}) suggests the following scheme:

\noindent (A) At sufficiently weak magnetic fields the differential
conductivity $G = {\rm d} I / {\rm d} V$ exhibits a single peak centered at
zero bias voltage (Fig.~\ref{Peaks}a).  The form of the peak is consistent
with a Lorentzian-type dependence of $I$ on $V$~\cite{Turner}:
\begin{equation}
            \frac{I}{V} = D \frac{2 \Gamma}{(e V)^2 + \Gamma^2},
\label{Lorentz}
\end{equation}
where $\Gamma$ is the tunneling peak width and $D$ is some constant.

\noindent (B) At larger magnetic fields the peak broadens and a small
depression in $G$ as a function of $V$ at $V = 0$ develops.  Thus, the
dependence $G(V)$ has two maxima (Fig. 10b).

\noindent (C) With further increase in the magnitude
of, both the total width of the
feature and the distance $E_{\rm tun}$ between the two maxima increases.
The latter distance appears to be linear in magnetic field,
\begin{equation}
                      E_{\rm tun} = 0.45 \hbar \omega_c,
\end{equation}
see Fig.~\ref{Peaks}c.

We are interested mainly in regime C where the magnetic field is not too
weak. Nevertheless, for the sake of completeness, we will discuss the other
two regimes (A and B) as well. We associate the existence of the three
different regimes (A, B, and C) with different relations among three energy
scales: $\hbar\omega_c$, $\gamma$, and $E_{\rm ex} \sim r_s \hbar\omega_c$.
They characterize the strength of the magnetic field, the LL broadening
due to the disorder, and the strength of the
electron-electron interactions, respectively.

%
%

\noindent (A) {\it Shubnikov-de Haas regime}. This regime corresponds to the
condition $\hbar\omega_c \ll \gamma$, where the dominating energy scale is
due to the disorder. In this case the r\^ole of the electron-electron
interactions reduces mainly to the screening of the impurity potential.
Denote the screened potential by $W(\bbox{r})$. The tunneling properties of
the system can be adequately analyzed within a simple model of
non-interacting electrons in zero magnetic field subjected to the external
potential $W(\bbox{r})$.  Such a theory leading to formula~(\ref{Lorentz})
was developed by Zheng and MacDonald~\cite{Zheng}. In the simplest case where
the two wells have identical densities, amounts of disorder, and where the
disorder in the wells is uncorrelated, the results of Zheng and MacDonald can
be understood from a well-known formula (see, e.g., Shrieffer {\it et
al.}~\cite{Shrieffer})
\begin{eqnarray}
&\displaystyle I = \frac{e}{\hbar^3} \sum_{\bbox{k} \bbox{p} \sigma}
|T_{\bbox{k} \bbox{p}}|^2 \int \frac{{\rm d} E}{2 \pi}
A_{\bbox{k} \sigma}^L(E) A_{\bbox{p} \sigma}^R(E + eV)&\nonumber\\
&\mbox{} \times [f(E) - f(E + eV)].&
\label{I A}
\end{eqnarray}
In this formula $T_{\bbox{k} \bbox{p}}$ is the tunneling matrix element,
$A_{\bbox{k} \sigma}(E)$ is the spectral density for energy $E$, momentum
$\hbar \bbox{k}$ and spin $\sigma$.  The superscripts $L$ and $R$ stand for
the left and right wells. Finally, $f(E)$ is the Fermi-Dirac distribution
function.  Equation~(\ref{I A}) shows that the tunneling experiments measure
the convolution of the wells' spectral functions.

The most important assumption for the derivation of Eq.~(\ref{Lorentz}) (and
for the resonant character of zero-bias tunneling in general) is that the
momentum is conserved during tunneling, e.g., that $|T_{\bbox{k} \bbox{p}}|^2
\propto \delta_{\bbox{k}, \bbox{p}}$. If now, following
Ref.~\onlinecite{Zheng}, one takes advantage of the Born approximation
expression
\begin{equation}
A_{\bbox{k} \sigma}(E) =
\frac{\hbar^2 / \tau}{(E - E_k)^2 + ({\hbar} / {2 \tau})^2}
\label{Born}
\end{equation}
for the spectral function, then one recovers Eq.~(\ref{Lorentz}) with $\Gamma
= \hbar / \tau$. Here, $\tau$ is the quantum lifetime and $E_k = \hbar^2 k^2
/ 2 m$. As noted above, Eq.~(\ref{Lorentz}) describes the experimental
results rather well~\cite{Turner,Zheng,Murphy}. However, before we proceed to
the case of stronger magnetic fields (regime B), we note that in dirtier
samples one should observe a different $I(V)$ dependence, owing to the fact
that the Born approximation breaks down. We will show that the disorder
broadening $\gamma$ of the LLs, and, consequently, the tunneling peak width,
can be much larger than $\hbar / \tau$.  To do so, we will need to make
several definitions first. In the studied samples the disorder is presumably
due to randomly positioned ionized donors.  In this case the correlation
length of the disorder potential is of the order of spacer width $d$, which
can easily be of the order of $1000{\rm\AA}$.  Let $n$ be the 2DEG density,
and $n_i$ be the density of randomly positioned donors. It is easy to
calculate then that the rms value of the screened disorder potential in the
plane of the 2DEG is equal to~\cite{Efros}
\begin{equation}
U \equiv \sqrt{\langle W^2 \rangle} =
\left(\frac{\pi}{8}\right)^{1/2} \frac{e^2 a_{\rm B} \sqrt{n_i}}{d}
\end{equation}
The Born approximation and, consequently, Eq.~(\ref{Lorentz}) are valid if $U
\ll \hbar v_{\rm F} / d$, which is the same as $n \gg n_i$. In the opposite
case, $n \ll n_i$, one has to use the quasiclassical approximation, which
leads to
\begin{equation}
A_{k \sigma}(E) = \frac{\sqrt{2 \pi}\,\hbar}{U}
\exp \left[-\frac{(E - E_k)^2}{2 U^2}\right]
\label{A Gauss}
\end{equation}
for the spectral function and
\begin{equation}
\frac{I}{V} = \frac{\sqrt{\pi}\, D}{U}
\exp\left[-\frac{(e V)^2}{4 U^2}\right]
\label{Gauss}
\end{equation}
for the tunneling current. Technologically, it is nowadays possible to change
both $n$ and $n_i$ in a given sample.  The former may be done by applying a
voltage to gates located nearby the quantum wells~\cite{Glozman}, and the
latter by special techniques of the sample's cool-down~\cite{Buks}. Hence, it
is possible to see the crossover from Eq.~(\ref{Lorentz}) to
Eq.~(\ref{Gauss}) experimentally. In this connection we mention a relation $U
= \frac{\hbar}{\tau} \sqrt\frac{n}{n_i}$, which ensures that the change in
the tunneling peak width from $\gamma = \frac{\hbar}{\tau}$ (at $n \gg n_i$)
to $\gamma = U$ (at $n \ll n_i$) is continuous~\cite{Comment on tails}.
Concluding the consideration of regime A, note that in terms of transport
measurements, it corresponds to the Shubnikov-de Haas effect,
which is reflected in the name of this regime~\cite{Comment on
SdH}.

Since the data of Ref.~\cite{Turner} appears to agree with
Eq.~(\ref{Lorentz}), corresponding to $n \gg n_i$, we will assume this
inequality to hold in the following. In this case, regime A can be (somewhat
arbitrarily) defined as $\omega_c\tau < \pi$.

\noindent (B) {\it Spin-unresolved QHE regime}. When the magnetic field is
increased, we switch from regime A to regime B, where $\gamma \ll
\hbar\omega_c$ yet $\gamma \gg E_{\rm ex}$. It corresponds to the
spin-unresolved QHE in transport measurements. The LLs with different $N$ are
now well-defined; however, the disorder is still strong enough to cause the
collapse of the spin-splitting of the LL spin subbands~\cite{Fogler}. Hence,
the ground state is not spin polarized, and the CDW at the upper LL does not
appear yet.

As in regime A, the interaction among electrons can be treated on the
mean-field level, and the main interaction effect is the screening of the
impurity potential. It can be shown that the screening is performed largely
by the electrons occupying the lower completely filled LLs and the screened
potential $W(\bbox{r})$ is changed little from itself in zero
field~\cite{Fogler}.

In regime B, the shape of the tunneling peak is determined by the convolution
of the upper LL DOS of the two wells. To show this, in Eq.~(\ref{I A}) we
switch from $(k_x, k_y)$ to $(X, n)$ (the guiding center coordinate and the
LL index) representation.  The basis states in this representation are given
by Eq.~(\ref{Stick}). At zero temperature the expression for the tunneling
current becomes
\begin{equation}
I = \frac{e}{\hbar^3} T^2 \sum_{X n \sigma} \int\limits_0^{e V}
\frac{{\rm d} E}{2 \pi} A_{X n \sigma}^L (E - e V)
A_{X n \sigma}^R(E).
\end{equation}
This equation allows two simplifications. First, at bias voltages $|eV| <
\hbar\omega_c$ we have to retain only the terms with $n = N$ in the sum.
Second, it is easy to understand that $A_{X N \sigma}(E)$ does not depend
neither on $\sigma$ nor on $X$. Instead of $A_{X N \sigma}(E)$, it is more
convenient to use another quantity, $g(E)$, which depends only on $E$:
\begin{equation}
       g(E) = \frac{1}{2 \pi \hbar} A_{X N \sigma}(E).
\end{equation}
Clearly, $g(E)$ is the DOS at the upper LL. In agreement with our statement
above we find that
\begin{equation}
              I = 2 \displaystyle \frac{\pi D \hbar\omega_c}{e}
\int\limits^{e V}_0 {\rm d} E g(E - e V) g(E),
\label{I B}
\end{equation}
i.e., that the tunneling current is determined by the convolution of the DOS
in the two wells. The factor of two in Eq.~(\ref{I B}) accounts for the spin
degeneracy. The constant $D$ is the same as in Eq.~(\ref{Lorentz}).

The characteristic width $\gamma$ of LLs has a square-root dependence on the
magnetic field~\cite{Raikh},
\begin{equation}
            \gamma = \sqrt{\hbar\omega_c \frac{\hbar}{2 \pi\tau}}
\label{gamma}
\end{equation}
Owing to the convolution, the width of the tunneling peak is by a factor of
two larger than $\gamma$. As for the shape of the tunneling peak, it depends
on the relation between the magnetic length $l$ and the correlation length
$d$ of the disorder potential~\cite{Comment on shape}.

There is still one more issue to address when considering regime B. In the
experiment, one can see a small depression in $G(V)$ in the vicinity of zero
bias (Fig.~\ref{Peaks}b). Such a depression can not be explained within the
model where the electron-electron interactions are treated on the mean-field
level. We think that the observed depression is a manifestation of a
correlation effect, namely, the Coulomb gap~\cite{Efros_Shklovskii}.  At
present, the theory of the Coulomb gap is developed only for strongly
localized, i.e., almost classical particles. The quantum-mechanical effects
have been studied numerically within the Hartree-Fock
approximation~\cite{Yang}. It is not clear whether or not the ideas of the
classical Coulomb gap are at all applicable to the system we are studying
now. If they do, a na\"{\i}ve estimate for the characteristic width of the
depression will be the energy of the Hartree interaction at the distance of
the order of the magnetic length $l$. This energy is larger than $E_{\rm h}$
but smaller than $E_{\rm ex}$. For a better estimate, a deeper
understanding of the Coulomb gap in the QHE regime is
required.

\noindent (C) {\it Spin-resolved QHE regime}. This regime is realized at even
stronger magnetic fields where $E_{\rm ex}$ becomes larger than
$\gamma$~\cite{Fogler}.  The electron-electron interactions are now the most
important, while the disorder can be treated as a weak perturbation.

Depending on how large the ratio $n / n_i$ is, the evolution of the
thermodynamical and transport properties of the system undergoes one or
multiple stages as the magnetic field increases. To avoid complexity, let us
consider only the case $n_i \ll n \ll n_i / \alpha^3$, where $\alpha = E_{\rm
ex} / \hbar\omega_c$.

The transition from regime B to regime C is associated with several dramatic
changes~\cite{Fogler}. First, the LLs become spin-split and in transport
measurements one should see spin-resolved conductivity peaks.  Second, the
nature of screening changes. Now it is performed mainly by the upper LL and
it is stronger than in zero magnetic field. As a result, the amplitude of the
random potential drops by a factor of the order of $\alpha$, and  $\gamma \to
\alpha\gamma$.  Thus, in this regime the disorder is additionally suppressed.
Third, at the upper spin subband the CDW appears. As a result, the DOS
acquires van Hove singularities separated by a pseudogap of the order of
$E_g$. As for the tunneling current, it is related to the DOS by
\begin{equation}
              I = \displaystyle \frac{\pi D \hbar\omega_c}{e}
\int\limits^{e V}_0 {\rm d} E g(E - e V) g(E),
\label{I C}
\end{equation}
which differs from Eq.~(\ref{I B}) by a factor of two due to the fact that
the spin degeneracy is lifted.  We will show below that at moderate $N$, the
differential conductance $G(V)$ exhibits two sharp maxima separated by a
pseudogap whose width we denote by $E_{\rm tun}$ (Fig.~\ref{Peaks}c).

Equation~(\ref{I C}) shows that the tunneling current is determined by the
DOS. In a disorder-free system the DOS is given by
\begin{equation}
        g(E) = \frac{2 \pi l^2}{L_x L_y} \sum \delta(\epsilon_i - E),
\label{DOS clean}
\end{equation}
$\epsilon_i$ being the energy levels in the self-consistent HF potential
$\epsilon(x, y)$. We will consider the stripe and ``bubble'' phases
separately.

{\bf Stripe phase.} In this case the energy levels are given by
$\epsilon_i = \epsilon(i a)$, $a$ defined by Eq.~(\ref{a}), which in the
limit $L_y \to \infty$ leads to
\begin{equation}
  g(E) = \frac{2}{\Lambda}\left\{
\begin{array}{cc}
\displaystyle\left|\frac{{\rm d}\epsilon}{{\rm d}x}\right|^{-1},&
|E| < \frac12 E_g\\
0, & |E| > \frac12 E_g.
\end{array}
\right.
\label{g 2D}
\end{equation}
We will start with the case of a theoretical interest, $N \gg 1$, where the
DOS is schematically shown in Fig.~\ref{DOS}. In the first approximation,
$g(E)$ vanishes at $|E| < \frac12 E_{\rm h}$ and $|E| > \frac12 E_g$, and so
does the tunneling current at bias voltages $|e V| < E_{\rm h}$ and $|e V| >
E_g$. More precisely, at $|e V| < E_{\rm h}$, the current first precipitously
drops towards $V = 0$ by a factor of the order of $N^2$ and then decreases
more slowly until at $V = 0$ it vanishes entirely.  Clearly, at large $N$ the
differential conductivity $G(V)$ at $|e V| < E_{\rm h}$ and $|e V| > E_g$ is
very small. Now examine the intermediate range of $e V$.

At $eV = \pm\frac12 (E_{\rm h} + E_g)$ and $eV = \pm E_g$, $G$ exhibits sharp
maxima associated with the presence of the delta-functions in $g(E)$. Thus,
$G$ has four maxima as a function of $V$, and the distance between the
furthermost ones is $2 E_g$.

Now let us see how these results are modified by a weak disorder.  Recall
that the impurity potential $W(\bbox{r})$ is strongly screened by the
electron gas. In general, it is necessary to know how such a screening is
achieved. For simplicity, we will discuss only the case of high magnetic
fields where $R_c \ll d$. In this case the screening is performed by
long-range fluctuations of the electron density at the upper LL.  We can say
that there is a random distribution of the local filling factor $\nu_N$,
described by the probability density $P(\nu_N)$. Clearly, in our model
$P(\nu_N)$ is close to the normal distribution
\begin{equation}
P(\nu_N) = \frac{1}{\sqrt{2 \pi \delta\nu}}
\exp\left[-\frac{(\nu_N  - \bar\nu_N)^2}{2 (\delta\nu)^2}\right].
\end{equation}
The characteristic spread $\delta\nu$ of such a distribution can be related
to $n_i$, in its turn related to the width of the tunneling peak $\Gamma =
\frac{\hbar}{\tau}$ in zero magnetic field:
\begin{equation}
(\delta\nu)^2 = \frac{n_i}{\sqrt{8 \pi} (n_{\rm L} d)^2}
= \frac{\sqrt{2}}{r_s}
\frac{(e^2 / \kappa d) (\hbar / \tau)}{(\hbar\omega_c)^2}.
\end{equation}
The DOS is given by
\begin{equation}
 g(E) = \int\limits_0^1 {\rm d} \nu_N P(\nu_N) g_0(\nu_N, E),
\label{g II}
\end{equation}
where $g_0(\nu_N, E)$ is the DOS in a disorder-free system for the filling
factor $\nu_N$. The energy $E$ is referenced with respect to the Fermi-level.

\begin{figure}
\vspace{0.1in}
\centerline{
\psfig{file=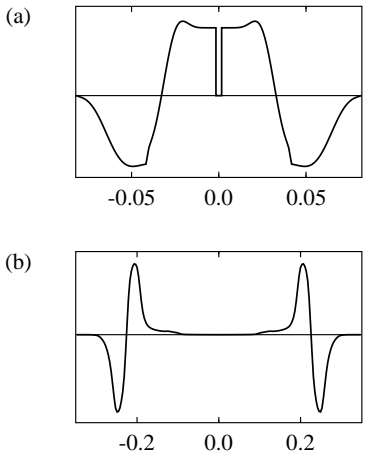,width=2.0in,bbllx=140pt,bblly=305pt,bburx=330pt,bbury=545pt}
}
\vspace{0.1in}
\setlength{\columnwidth}{3.2in}
\centerline{\caption{The differential
conductivity $G$ as a function of the
bias voltage $V$. $e V$
(horizontal axes) is in units of $\hbar\omega_c$.
(a) $N \gg 1$ (b) $N = 3$.
The other parameters used in generating these graphs,
are $r_s = 1$, $\bar\nu_N = \frac12$, and $\delta\nu = 0.05$. The
calculations are made for the stripe phase.
\label{Conductivity}
}}
\end{figure}

It is clear that the disorder causes smearing of all sharp features in the
DOS. Consequently, the main effect of the disorder is the removal of the
singularities in $g(E)$ at the extremes of the spectrum. In fact, these
singularities have a very small weight (it is equal to the fraction of the
CDW period where $\epsilon(x) = \text{const}$, see Fig.~\ref{Plot real}). As
a result of such a smearing, the sharp maxima in $G(V)$ disappear. The
remaining feature has the total spread of $2 E_g$ and it is as follows
(Fig.~\ref{Conductivity}a). At $|e V| < E_{\rm h}$, $G(V) \approx 0$; at
$E_{\rm h} < |e V| < \frac12 (E_{\rm h} + E_g)$, the differential
conductivity is positive and approximately constant.  At $|e V| \approx
\frac12 (E_{\rm h} + E_g)$, $G(V)$ sharply drops, crosses zero and becomes
equal to a negative constant with about the same absolute value.  Finally, at
$|e V| = E_g$, $G(V)$ rapidly rises to reach zero.  In the limit of large
$N$, $E_g \gg E_{\rm h}$, and so $G(V)$ does not exhibit any sharp maxima.
The small ``bumps'' visible in Fig.~\ref{Conductivity}a are the only
reminders of the distinct van Hove singularities of the clean case.  Clearly,
in the dirty case our definition of the tunneling pseudogap becomes very
unnatural.  It is more logical to associate the pseudogap with the range
$-E_{\rm h} < e V < E_{\rm h}$ where $G(V)$ is small. As for the larger
energy scale $E_g$, it describes the total width of the tunneling peak.

%
%

The described picture applies to the case of large $N$.  However,
experimentally accessible $N$ ($1 \leq N \leq 4$) are not large. For such $N$
we calculated $G(V)$ numerically. As an example, the results for $N = 3$ are
shown in Fig.~\ref{Conductivity}b where one can see two sharp maxima in
$G(V)$. We found that $E_{\rm tun}$ can be satisfactorily fitted to a linear
law $E_{\rm tun} \approx 0.4 \hbar\omega_c$.  We would like to remind that
this expression is obtained for $\bar\nu_N = \frac12$. In general, $E_{\rm
tun}$ more or less linearly decreases as $\bar\nu_N$ becomes smaller.  At
$\bar\nu_N \lesssim 1 / N$ this dependence becomes sublinear. Eventually,
$E_{\rm tun}$ vanishes altogether at $\bar\nu_N = 0$.

{\bf ``Bubble'' phase.} In this case to obtain the DOS [Eq.~(\ref{DOS
clean})] one has to find the self-consistent HF potential $\epsilon(x, y)$
and then solve the Schr\"odinger equation with this potential to find the
energy levels $\epsilon_i$. As in any periodic system, they will form energy
bands. Certainly, the stripe phase we studied above is also periodic, and,
therefore, its DOS also has a band structure. However, in the case of the
stripe pattern there is only one (partially occupied) band, so the mentioning
of the band structure would have been superfluous. On the contrary, for the
``bubble'' phase there are several bands, and their total number depends on
the number of flux quanta penetrating a unit cell of the lattice formed by
the ``bubbles''. When this number is not a simple rational fraction, the
number of bands is quite large.  However, many of such bands may be
almost degenerate. A simple tight-binding-like picture can be developed for
large $N$. In this case the band widths are exponentially small. Roughly
speaking, their width is proportional to the overlap of two coherent states
separated by the period of the ``bubble'' lattice, i.e, the distance of the
order $3 R_c \gg l$. Hence, at large $N$ the DOS can be approximated by a
set of narrow peaks. Recall that every ``bubble'' resides in a potential well
created by the HF interaction. It follows from the quasiclassical
consideration that when the number of electrons $n_e$ in one ``bubble'' is
large, the positions of the peaks in the DOS are simply the energies of the
constant energy contours in the self-consistent potential $\epsilon(x, y)$,
enclosing an integer number of flux quanta in this potential well. Thus, our
predictions for the tunneling experiments are as follows. In a sufficiently
clean sample, the differential conductivity exhibits many peaks. The distance
in energy between the furthermost peaks is of the order of $E_g$ (as in the
case of the stripe pattern). Also, similar to the case of stripes, there
exists a gap of width $2 E_{\rm h}$ centered at zero bias. However, unlike in
the case of stripes, the current is not just strongly suppressed at $|e V| <
E_{\rm h}$, but vanishes exactly because in this case this is a true gap, not
a pseudogap.  This concludes our analysis of regime C.

Previously, the conclusion about the existence of the tunneling gap with the
width $2 E_{\rm h}$ was reached in Refs.~\onlinecite{Aleiner,Aleiner_hydro}.
However, their predictions for the overall shape of the DOS are different
from ours. Unlike the broad feature with the overall width $E_g \gg E_{\rm
h}$ we derive, their results are that the DOS consists of just two narrow
peaks. Thus, the energy scale $E_g$ does not appear in their DOS.

Recently, Levitov and Shytov~\cite{Levitov} also argued that the tunneling
conductance is represented by two narrow peaks.  Let us again use $E_{\rm
tun}$ to denote the distance between the peaks in $G(V)$. In our notations,
the result of Levitov and Shytov is
\begin{equation}
                          E_{\rm tun} = 2 E_{\rm ex}.
\label{Levitov}
\end{equation}
By $E_{\rm ex}$ in this formula, we mean the exchange gap for the spin
excitations in one well for the double-well system. It differs from the one
given by Eq.~(\ref{E_ex}), derived for an isolated well, by arguments of the
logarithms. [This originates from the difference in the screening properties
of a single-well and of the double-well system (see below)]. As for a
single-well system, $E_{\rm ex}$ is linear in field at large $N$.
Formula~(\ref{Levitov}) was derived in Ref.~\onlinecite{Levitov} under an
assumption that the dynamics of the system can be described by fluctuations
of the Fermi surface. Apparently, in this approach the discreteness of the
LLs is lost. It is not accidental then that $E_{\rm tun}$, as given by
Eq.~(\ref{Levitov}), does not depend on the filling $\bar\nu_N$ of the upper
LL. At the same time, it is clear that such dependence does exist. Indeed,
consider $\bar\nu_N \ll 1 / N$ case where the ground state is a dilute WC.
This system can be treated semiclassically with the result that the tunneling
gap is equal to twice the energy difference of a vacancy and an interstitial,
which is of the order of the Hartree interaction on the distance between
nearest neighbors. This energy is much smaller than $E_{\rm ex}$, and,
moreover, vanishes altogether in the limit $\bar\nu_N \to 0$.  More
generally, it can be shown that $E_{\rm tun}$ never exceeds $2 E_{\rm ex}$,
and Eq.~(\ref{Levitov}) holds only in the limit $\bar\nu_N \to +0$ in one
well, while $\bar\nu_N \to 1 - 0$ in the other. It does not hold in the case
of equal densities, it was proposed for.

Concluding this Section, we briefly discuss the proximity effects, important
if the separation $b$ between the two wells is comparable with $a_{\rm B}
\sim 100{\rm\AA}$. If we go again through the derivation of Eq.~(\ref{I C}),
keeping in mind that we are examining regime C now, it is easy to realize
that we, in fact, assumed that any reasonable amount of the disorder would be
sufficient for the self-averaging of $A_{N X}$ in the sample and that the
phases of the CDW in the two wells are uncorrelated. We also ignored the
interaction between the tunneling electron and the hole it leaves
behind~\cite{Eisenstein_95}. Let us examine how such effects can modify our
results.

Ideally, when the two wells are brought close together, the CDW patterns
existing in each well should lock in the anti-phase to reduce the Hartree
energy of the system. It can be understood on the example of the limiting
case of the two wells located just next to each other. In this case the
Hartree energy is reduced to zero because the charge oscillations in one well
are compensated by the charge oscillations in the other, and the total charge
no longer oscillates. This effective suppression of the Hartree potential
would lead to an increase in the optimal CDW period.  Correspondingly, in the
expression~(\ref{E_g II}) for $E_g$ one has to use larger $\Lambda$, i.e.,
$E_g$ tends to increase.  However, the phase locking energy, or the
difference in energy for the anti-phase and in-phase arrangements turns out
to be small for the parameters of Ref.~\onlinecite{Turner}. We expect
that this phase locking effect is totally washed out at experimentally
accessible temperatures and amounts of disorder.  Therefore, it is more
reasonable to assume that the CDW in the two wells are uncorrelated as we did
above.

More important effect is the enhancement of the dielectric constant.
It can be shown that for the double-well system the single-well dielectric
constant $\varepsilon(q)$ [Eq.~(\ref{epsilon})] gets replaced by a larger
value of
\begin{equation}
  \varepsilon(q) \to \varepsilon(q)\frac
     {{\rm e}^{2 q b} - [1 - \varepsilon(q)^{-1}]^2}
     {{\rm e}^{2 q b} - [1 - \varepsilon(q)^{-1}]}.
\end{equation}
For example, at $b = 0$ the dielectric constant roughly doubles. The
stronger screening leads to decrease in $E_{\rm tun}$.

Finally, there is also a so-called excitonic shift, which accounts for the
interaction between the negatively charged tunnelling electron and the
positively charged hole it leaves behind. The excitonic shift reduces $E_{\rm
tun}$ as well. For example, in the limiting case $b = 0$, $E_{\rm tun}$
vanishes altogether. In practice, however, one has the inequality $a_{\rm B}
\ll b \ll R_c$, and the aforementioned effects cause a small correction to
$E_{\rm tun}$. Our estimate of such a correction is
as follows:
\begin{eqnarray}
E_{\rm tun}(b) &=& E_{\rm tun}(\infty) - \hbar\omega_c\nonumber\\
\displaystyle &\times& \left[
\frac{\ln\left(2 N r_s \sqrt{a_{\rm B} / b}\right)}{N} +
\frac{\pi r_s}{12\sqrt{2}} \frac{a_{\rm B}^2}{b^2}\right].
\label{Exciton}
\end{eqnarray}
%
\section{Conclusion}
\label{Conclusion}

In this paper we showed that, in the framework of the Hartree-Fock
approximation, the ground state of 2D electron gas in a weak magnetic field
is a CDW at the upper partially filled LL.  Both the cohesive energy per
electron at the upper LL and the characteristic width of the LLs have the
scale of the exchange-enhanced spin-splitting of the upper LL. This energy
is smaller than $\hbar\omega_c$ for $r_s \lesssim 1$, and thus, the LLs are
not destroyed by the electron-electron interaction.

As the magnetic field decreases, the fraction of electrons, participating
in the CDW, goes to zero, so that at zero magnetic field the density
is uniform.

The existence of the CDW leads to the pseudogap in the one-particle DOS
centered at the Fermi energy. The calculated width of the pseudogap seems to
be in a good agreement with the width of the pseudogap observed in the
tunneling conductance of the double-well system~\cite{Turner}.

The CDW at the upper LL strongly affects the low-temperature transport
properties of the 2D gas. Due to the pinning of the CDW by disorder, the
dissipative conductivity $\sigma_{\rm xx}$ has narrow peaks at half-integer
fillings even in high-mobility heterostructures.
At higher temperatures the depinning of the CDW becomes possible.  The effect
of this phenomenon on the transport properties remains to be studied. At the
moment, we can only estimate the temperature $T_c$ at which the CDW melts into
a perfectly uniform electron liquid. In the spirit of Ref.~\cite{Fukuyama},
this estimate is
\begin{equation}
    k_B T_c = \bar\nu_N ( 1 - \bar\nu_N ) \tilde{u}_{\rm ex}(q_0).
\end{equation}
At $r_s \sim 1$, we can use the asymptotical formula~(\ref{u_ex
asym}) to get
\begin{equation}
 k_B T_c(\bar\nu_N = \frac12) \approx 0.02 \hbar\omega_c + 0.06 E_{\rm h}.
\label{T_c}
\end{equation}
Above $T_c$, the pinning effects disappear completely; the peaks in
$\sigma_{\rm xx}$ become wide; the plateaus in $\sigma_{\rm xy}$ become
narrow. At $N = 5$, Eq.~(\ref{T_c}) gives $T_c \approx 0.03 \hbar\omega_c$ in
reasonable agreement with experimental data~\cite{Stormer,Rokhinson}.
\acknowledgments

Useful discussions with I.~L.~Aleiner, L.~I.~Glazman, D.~R.~Nelson, and
I.~M.~Ruzin are greatly appreciated. We are grateful to V.~J.~Goldman and
L.~P.~Rokhinson for communicating to us their unpublished results. This work
was supported by NSF under Grant No.~DMR-9321417.
\appendix
\section{Derivation of Eq.~(\ref{B})}
\label{Derivation of B}

Using Eqs.~(\ref{epsilon},\ref{u_H},\ref{u_ex},\ref{u_HF 1D}), we find
\begin{eqnarray}
&u_{\rm ex}(x) = \displaystyle\frac{\hbar\omega_c}{L_y}
\int\limits_0^\infty\!\displaystyle\frac{{\rm d} y\, F^2(r / l^2)}
{1 + \frac{r}{\xi} - J_0^2(r k_{\rm F})},&
\label{u_ex x}\\
&r \equiv \sqrt{x^2 + y^2}.&
\end{eqnarray}
We will use the following asymptotic formula for $F(q)$, which can be derived
by the saddle-point integration in Eq.~(\ref{F def}), using the WKB
approximation for the wave-functions~(\ref{Stick}) in the integrand:
\begin{equation}
F(q) \simeq \sqrt{\frac{2}{\pi q s R_c}}\,
\cos\left[\frac{q s R_c}{2} +
k_{\rm F} R_c \arcsin\!\left(\frac{q}{2 k_{\rm F}}\right) -
\frac{\pi}{4}\right],
\end{equation}
where
\begin{equation}
                  s \equiv \sqrt{1 - (q / 2 k_{\rm F})^2}.
\end{equation}
This formula is valid for $q R_c,\,\,(2 k_{\rm F} - q) R_c \gg N^{2 / 3}$.
Note that it agrees with Eq.~(\ref{F asym}) but has a broader region of
validity towards large $q$.

By means of this formula, Eq.~(\ref{u_ex x}) can be transformed into
\begin{equation}
\displaystyle u_{\rm ex}(x) \simeq \frac{a \hbar\omega_c}{\pi^2 R_c}
\!\!\!\int\limits_0^{\sqrt{4 R_c^2 - x^2}}\!\!\!\!\!\!
\frac{{\rm d} y}{r \sqrt{4 R_c^2 - r^2}\,
\left(1 + \displaystyle\frac{r}{\xi}\right)} + a\delta(x) E_{\rm h},
\end{equation}
which can be rewritten as
\begin{eqnarray}
&\displaystyle u_{\rm ex}^{\rm eff}(x) \simeq a\delta(x) E_{\rm h} +
\frac{a \hbar\omega_c}{\pi^2}
\int\limits_0^{\sqrt{4 R_c^2 - x^2}}\!\!\!\!\!\!
\frac{{\rm d} y}{\sqrt{4 R_c^2 - x^2 - y^2}}&\nonumber\\
&\displaystyle \mbox{} \times \left(\frac{1}{\sqrt{x^2 + y^2}} -
\frac{1}{\xi + \sqrt{x^2 + y^2}}\right).&
\end{eqnarray}
On the other hand, the Hartree part for $a_{\rm B} < |x| < c R_c$, $c$ being
a number of order unity, can be approximated by
\begin{equation}
u_{\rm H}^{\rm eff}(x) = \frac{2}{L_y}\!\!\!
\int\limits_0^{\sqrt{4 R_c^2 - x^2}}\!\!\!\!\!
{\rm d}y \frac{e^2 a_{\rm B}}{\pi r \kappa\sqrt{4 R_c^2 - x^2 - y^2}}
+ \text{const},
\end{equation}
see Eq.~(\ref{u_H r asym}) below. Combining the last two equations, we
obtain Eq.~(\ref{B}).
\section{Wigner crystal revisited}
\label{WC}

In this appendix we calculate the cohesive energy $E_{\rm coh}^{\rm WC}$ of
the HF WC state and then compare it with the cohesive energy $E_{\rm
coh}^{\rm CDW}$ of the CDW state. As was explained in Sec.~\ref{CDW}, our
CDW state differs from the WC at $1 / (N r_s^2) \lesssim \bar{\nu}_N \leq
\frac12$. We will show that in this entire range $E_{\rm coh}^{\rm CDW} <
E_{\rm coh}^{\rm WC}$, i.e., the CDW state is indeed more energetically
favorable. The qualitative arguments in favor of this statement were given in
Sec.~\ref{Quality}.

Earlier, the cohesive energy $E_{\rm coh}^{\rm WC}$ of the WC has been
calculated in Ref.~\onlinecite{Aleiner}. For $\bar{\nu}_N < 1 / N$ our
results agree with Ref.~\onlinecite{Aleiner}. At larger $\bar{\nu}_N$, they
differ.

To calculate $E_{\rm coh}^{\rm WC}$ we have to find the set of
${\Delta}(\bbox{q})$ corresponding to the WC state, and then substitute them
into the general formula~(\ref{E_coh}).  Using Eq.~(\ref{Delta coherent}),
the additivity of order parameter (see Sec.~\ref{num res}), and the Poisson
summation formula, it is easy to obtain that non-zero ${\Delta}(\bbox{q})$
correspond to the wave-vectors of the reciprocal lattice of the WC
\begin{equation}
\bbox{q}_{i, j} = Q_0
\left( i + \frac12 j,\ \frac{\sqrt3}{2} j \right),\quad
Q_0 = \sqrt{ \frac{4\pi \bar{\nu}_N} {\sqrt3} }\,\frac{1}{l},
\label{rec_lattice}
\end{equation}
for which $q$ they are given by
\begin{equation}
        \Delta(q) = {\bar{\nu}_N} \exp\left(-\frac14 q^2 l^2\right),
\label{Delta WC}
\end{equation}
derived earlier in Ref.~\onlinecite{Yoshioka} for the WC at the lowest LL.
Hence, for the cohesive energy of the WC we obtain
\begin{equation}
E_{\rm coh}^{\rm WC} = \frac{{\bar{\nu}_N} n_{\rm L}}{2}
\!\!\sum_{|i| + |j| > 0}
\!\!\!\tilde{u}_{\rm HF}\left(\bbox q_{i, j} \right)
\exp\left( -\frac{1}{2} \bbox{q}_{i, j}^2 l^2 \right).
\label{E_coh WC}
\end{equation}
%

\noindent{$\bbox{\bar{\nu}_N = 1/2}$.}
At $\bar{\nu}_N = 1/2$ the contribution of the six shortest reciprocal
lattice vectors $(i, j) \in \{ (\pm 1, 0),\: (0, \pm 1),\: (1, -1),\: (-1,
1)\}$ constitutes more than $97\%$ of the sum~(\ref{E_coh WC}).  Therefore,
with a good accuracy one can write
\begin{eqnarray}
E_{\rm coh}^{\rm WC} &=& \displaystyle\frac32 n_{\rm L}
\tilde{u}_{\rm HF}(Q_0) {\rm e}^{-\pi / \sqrt{3}},
\label{E_coh WC I}\\
Q_0 &=& \displaystyle\sqrt{ \frac{2 \pi}{\sqrt{3}} }\, \frac{1}{l}.
\end{eqnarray}
Analyzing this expression, we discover a remarkable fact that $E^{\rm
WC}_{\rm coh}$ can be positive. In other words, the WC loses competition even
to the uniform state (UEL)! Indeed, consider the limit $N \gg r_s^{-2} \gg
1$. It follows from Eqs.~(\ref{u_H asym},\ref{u_ex asym}) that at $q \sim Q_0
\sim l^{-1}$ the HF potential is dominated by the (non-negative) Hartree
potential $\tilde{u}_{\rm H}(q)$ exhibiting oscillations in $q$. Roughly, the
oscillating part of $\tilde{u}_{\rm H}(q)$ is proportional to $\sin (2 q
R_c)$. Since different $N$ correspond to different values of $\sin (2 Q_0
R_c)$, $E^{\rm WC}_{\rm coh}$ oscillates as well. For example, if $\sin (2
Q_0 R_c) = -1$, or, more precisely, if $Q_0$ is one of the zeros of
form-factor $F(q)$, the Hartree term $\tilde{u}_{\rm H}(Q_0)$ is also zero.
In this case $E^{\rm WC}_{\rm coh}$ has a minimum and its value is negative.
On the other hand, if $Q_0$ coincides with a maximum of $F(q)$, then the HF
potential and, consequently, $E^{\rm WC}_{\rm coh}$ are both positive.

The quantity $\sin (2 Q_0 R_c)$ is a pseudo-random function of $N$. Its
average value within an interval $N \in (r_s^{-2}, N_{\rm max})$ tends to
zero as $N_{\rm max} \to \infty$. Hence, roughly at every other $N$ the
cohesive energy $E_{\rm coh}^{\rm WC}$ is positive and the WC (at least, with
the triangular lattice) at $\bar{\nu}_N = \frac12$ is absolutely unstable.

Nevertheless, there exist $N$ at which $E^{\rm WC}_{\rm coh}$ is negative,
and the lower bound for $E^{\rm WC}_{\rm coh}$ can be found assuming that
$Q_0$ coincides with a zero of the Hartree potential. In this case
$\tilde{u}_{\rm HF}(Q_0) = -\tilde{u}_{\rm ex}(Q_0)$.  Using Eqs.~(\ref{u_ex
asym},\ref{E_coh WC I}), one obtains
\begin{equation}
\min\{ E_{\rm coh}^{\rm WC} \} \approx -
0.01 \frac{\hbar\omega_c}{\sqrt{N}} - \frac{E_{\rm h}}{4}.
\label{E_coh WC 1_2}
\end{equation}
This result should be compared with the energy of the CDW state given by
Eq.~(\ref{E_coh CDW}). The last term in the two formulas is identical.  In
fact, it is common for any low-energy state at $N$-th LL at $\bar{\nu}_N
\gtrsim 1 / N$ (see Sec.~\ref{Quality}). Therefore, we have to compare only
the remaining terms. These other terms are negative in both $E_{\rm coh}^{\rm
WC}$ and $E_{\rm coh}^{\rm CDW}$. In the limit $N \gg r_s^{-2}$, we are
considering now, the absolute value of the term for the CDW state is much
larger than of that for the WC state. Hence, the CDW state is more
energetically favorable. In addition to the analytical arguments, we also
compared $E_{\rm coh}^{\rm WC}$ and $E_{\rm coh}^{\rm CDW}$ numerically.
We found that $E_{\rm coh}^{\rm CDW}$ is always
smaller than $E_{\rm coh}^{\rm WC}$ in the parameter
range of interest: $1 \leq N \leq 10$ and $0.14 < r_s < 1.2$. However, at
such, rather moderate $N$, the difference between the two is not too large (of
the order of $5\%$).

\noindent$\bbox{\bar{\nu}_N \ll 1/2}$.
In this subsection we will consider only the limit $N \gg r_s^{-2}\gg 1$. In
principle, we can continue using the general formula~(\ref{E_coh WC}).
However, to give our analysis a new angle, we choose to use
formula~(\ref{E_coh num}), which, in the case of the WC, is nothing else than
the standard lattice sum~\onlinecite{Maki,Aleiner}:
\begin{equation}
 E_{\rm coh}^{\rm WC} = \frac12\sum_{\bbox{R}_i \neq 0} g_{\rm HF}(\bbox{R}_i)
- \frac{\bar{\nu}_N n_{\rm L}}{2} \int\! {\rm\bf d}^2 \bbox{r}
  g_{\rm HF}(\bbox{r}).
\label{E_coh WC real}
\end{equation}
The conventional estimate of such a lattice sum is
\begin{equation}
E_{\rm coh}^{\rm WC} \sim -\frac{\bar{\nu}_N n_{\rm L}}{2}
\int\limits_{\rm WS}\! {\rm\bf d}^2 \bbox{r} g_{\rm HF}(\bbox{r})
\sim -\frac12 g_{\rm HF}(a_0),
\label{Wigner-Seitz}
\end{equation}
where the integration is performed over the area of the Wigner-Seitz cell.
This is how the results of Ref.~\onlinecite{Aleiner} have been obtained.  As
long as the interaction potential $g_{\rm HF}(r)$ is sufficiently smooth at
$r \geq a_0$, $a_0$ being the lattice constant, this procedure is correct.
It can be shown, however, that at large $N$ the potential $g_{\rm HF}(r)$ is
given by [compare with Eq.~(\ref{u_HF asym r})]
\begin{equation}
 g_{\rm HF}(r) \sim \frac{e^2 a_{\rm B}}{\pi \kappa r \sqrt{4 R_c^2 - r^2}}
 + E_{\rm h},\quad r \gtrsim \xi,\: 2 R_c - r \gtrsim l,
\label{G_HF ASYM}
\end{equation}
i.e., it has a sharp maximum at $r = 2 R_c$ corresponding to the separation
at which the cyclotron orbits of the two states start intersecting. It is
due to this sharp maximum the Fourier transform of the HF potential
oscillates with the period $1 / (2 R_c)$.

Let us derive Eq.~(\ref{G_HF ASYM}). First of all, note that $\tilde{g}_{\rm
HF}(q)$ deviates from $\tilde{u}_{\rm HF}(q)$ only at $q \gtrsim l^{-1}$
[Eq.~(\ref{g_HF})]; hence, in the real space these two potentials essentially
coincide at $r \gtrsim l$.  Secondly, the exchange potential $u_{\rm ex}(r)$
rapidly decays at distances $r$ larger than $\xi = \sqrt{2}\, r_s R_c$ (see
the Sec.~\ref{CDW}) and soon becomes much smaller than the Hartree
potential; therefore,
\begin{equation}
               g_{\rm HF}(r) \sim u_{\rm H}(r),\quad r \gtrsim \xi.
\label{g_HF vs u_H}
\end{equation}
According to Eqs.~(\ref{u_H}),
\begin{equation}
  u_{\rm H}(r) = \int\!\frac{{\rm\bf d}^2 \bbox{q}}{(2 \pi)^2}
{\rm e}^{{\rm i}\bbox{q} \bbox{r}}\tilde{v}(\bbox{q}) F^2(\bbox{q}).
\end{equation}
At $\bbox{q}$ yielding the dominant contribution to this integral,
$\tilde{v}(\bbox{q})$ can be replaced by $\pi e^2 a_{\rm B} + (2 \pi)^2
E_{\rm h} \delta(\bbox{q})$ [cf. Eq.~(\ref{v r})]. Using also Eq.~(\ref{F
asym}) for $F(\bbox{q})$, we arrive at
\begin{eqnarray}
u_{\rm H}(r) &\approx& \pi e^2 a_{\rm B}\int\!
\frac{{\rm\bf d}^2 \bbox{q}}{(2 \pi)^2}
{\rm e}^{{\rm i}\bbox{q} \bbox{r}} J_0^2(q R_c) + E_{\rm h}\nonumber\\
&=& \frac{e^2 a_{\rm B}}{\pi \kappa r \sqrt{4 R_c^2 - r^2}} + E_{\rm h},
\label{u_H r asym}
\end{eqnarray}
which together with Eq.~(\ref{g_HF vs u_H}) leads to Eq.~(\ref{G_HF ASYM}).

Note that $u_{\rm H}(r)$ satisfies the relation
\begin{equation}
            u_{\rm H}(r) = n_{\rm L} \tilde{u}_{\rm ex}(r / l^2),
\label{u_H r}
\end{equation}
which follows from Eq.~(\ref{u_ex}). One may wonder why $u_{\rm ex}(q)$ as
given by Eq.~(\ref{u_ex asym}) does not show the inverse square-root
singularity present in $u_{\rm H}(r)$ according to Eq.~(\ref{u_H r asym}). It
is easy to see, though, that the singularity in $u_{\rm ex}(q)$ is located at
$q = 2 k_{\rm F}$, which is beyond the limited range of $q$, for which
Eq.~(\ref{u_ex asym}) is written.  Using the relation $\hbar\omega_c = e^2
a_{\rm B} / \kappa l^2$, one can verify that Eq.~(\ref{u_H r asym}) does
agree with Eq.~(\ref{u_ex asym}) in the indicated range of $q$.

Let us now return to the calculation of the cohesive energy of the WC.
Depending on the lattice constant $a_0$, one can distinguish two
possibilities: $a_0 \gg 2 R_c$ ($\bar{\nu}_N \ll 1/N$) and $a_0 \ll 2 R_c$
($1/N \ll \bar{\nu}_N \ll \frac12$). In the former case the singular part of
$g_{\rm HF}(r)$ has no effect and $E_{\rm coh}^{\rm WC}$ can be estimated
with the help of Eq.~(\ref{Wigner-Seitz}). In the latter case this standard
procedure fails. This case requires a more accurate treatment of the lattice
sites $\bbox{R}_i$ in Eq.~(\ref{E_coh WC real}) enclosed by the circle of
radius $2 R_c$. This can be done as follows.

We divide the entire area of the circle into narrow concentric rings of width
$\delta r$ and then sum the contributions to $E_{\rm coh}^{\rm WC}$ from all
rings. Denote by $M(r)$ the number of lattice sites in the ring with inner
radius $r$ and outer radius $r + \delta r$. Generally, $M(r)$ is a
pseudo-random function of $r$ with the average value of $\overline{M}(r) = 2
\pi r \delta r\bar{\nu}_N n_{\rm L}$.  Clearly, the contribution of the ring
under consideration to $E_{\rm coh}^{\rm WC}$ is $\frac12 g_{\rm HF}(r) [M(r)
- \overline{M}(r)]$. The largest contribution comes from rings with $r \sim 2
R_c$ where $g_{\rm HF}(r)$ has the maximum. In order that our procedure to make
sense the rings must contain at least one lattice site on average. On the
other hand, the accuracy of the estimate is higher if the rings are as narrow
as possible. Hence, the width of the rings $\delta r$ has to be determined
from the condition $\overline{M}(2 R_c) \sim 1$, which yields
\begin{equation}
                      \delta r \sim a_0^2 / (4 \pi R_c).
\label{delta r}
\end{equation}

Since $M(r)$ is pseudo-random, and the total number of rings $\sim R_c /
\delta r$ is finite, $E_{\rm coh}^{\rm WC}$ is also a pseudo-random quantity.
In other words, for any given $\bar{\nu}_N$, the cohesive energy $E_{\rm
coh}^{\rm WC}$ experiences fluctuations as a function of the ratio $R_c /
a_0$, i.e., as a function of $N$. This is exactly the conclusion we came to
in the preceding subsection by using different arguments. In can be verified
that the rms value of $E_{\rm coh}^{\rm WC}$ fluctuations is much larger than
its average value given by Eq.~(\ref{Wigner-Seitz}); therefore, $\min E_{\rm
coh}^{\rm WC}$ can be estimated as rms value of $E_{\rm coh}^{\rm WC}$ taken
with the negative sign.  To proceed we need to know the statistical
properties of the pseudo-random quantity $\delta M(r) \equiv M(r) -
\overline{M}(r)$. This is an interesting mathematical problem in itself. A
related problem, namely, the fluctuations in the number of the square lattice
sites contained in a circle as a function of its radius, was studied by many
mathematicians starting from the last century~\cite{Dyson}. The complete
solution is not yet obtained.  However, extensive numerical data indicate
that, provided $\delta r < a_0$, (i) rms value of $\delta M(r)$ is of order
$\overline{M}(r)$ and (ii) the fluctuations in $\delta M(r)$ and $\delta M(r
+ \Delta r)$ can be considered to be statistically independent if $\Delta r <
a_0$. In other words, the distribution of the lattice sites within any ring
of width $a_0$ resembles the completely random Poisson distribution. However,
the fluctuations in any two such rings are correlated. Clearly, the ring $2
R_c - a_0 < r < 2 R_c$ gives the dominating contribution to the fluctuations
in $E_{\rm coh}^{\rm WC}$ because in this ring $g_{\rm HF}(r)$ reaches its
maximum. This leads to the estimate
\begin{eqnarray}
(\min E_{\rm coh}^{\rm WC})^2 &\sim& \sum_{j = 0}^{a_0 / \delta r}
g_{HF}^2(r)\,\overline{M}(2 R_c - j \delta r)\nonumber\\
&\sim & \frac{\overline{M}(2 R_c)}{\delta r}
\int\limits_{\delta r}^{a_0}\! {\rm d} r\,
g_{HF}^2(2 R_c - r).
\end{eqnarray}
Evaluating this integral with the help of Eq.~(\ref{G_HF ASYM}), we get
\begin{equation}
\min E_{\rm coh}^{\rm WC}\sim
-\hbar\omega_c \left(\frac{\bar{\nu}_N} N\right)^{1/2} \! \ln^{1/2}\!\left[
\frac{a_0}{\max \left( \delta r,\: l\right) }\right].
\label{ecoh_1case}
\end{equation}
Remind that we are considering the case $1 / N \ll \bar{\nu}_N \ll \frac12$.
As one can see, at the upper limit it agrees (up to a numerical factor) with
the obtained earlier result for $\bar{\nu}_N = \frac12$ [Eq.~(\ref{E_coh WC
1_2})].

Finally, the results for different $\bar{\nu}_N$ may be summarized in the
following way:
\begin{equation}
\frac{\left|E_{\rm coh}^{\rm WC}\right|}{\hbar\omega_c} \lesssim \left\{\!
\begin{array}{cl}
r_s\sqrt{\bar{\nu}_N N},
&\displaystyle\bar{\nu}_N \ll \frac{1}{N^3 r_s^2} \\
\displaystyle\frac{\ln \left( \bar{\nu}_N N^3 r_s^2 \right)}{N},
&\displaystyle\frac{1}{N^3 r_s^2} \ll \bar{\nu}_N \ll \frac 1N \\
\displaystyle\left(\frac{\bar{\nu}_N}{N}\right)^{1/2}
\!\ln^{1/2}\left(\bar{\nu}_N N\right),
&\displaystyle\frac{1}{N} \ll \bar{\nu}_N \ll \frac 1{\sqrt{N}} \\
\displaystyle\left(\frac{\bar{\nu}_N}N\right)^{1/2}
\!\ln^{1/2} \!\left(\frac{1}{\bar{\nu}_N}\right),
&\displaystyle\frac{1}{\sqrt{N}} \ll \bar{\nu}_N \ll \frac12.
\end{array}
\right.
\label{E_coh WC all}
\end{equation}

At this point we can compare the energies of the WC and the CDW
[Eqs.~(\ref{E_coh CDW III}) and~(\ref{E_coh WC all}), respectively] at $1 /
(N r_s^2) \ll \bar{\nu}_N \ll \frac12$ where Eq.~(\ref{E_coh CDW III}) holds.
We see that the CDW state wins over the WC in this entire interval.

\end{multicols}

\begin{references}

\bibitem{Fukuyama}H.~Fukuyama, P.~M.~Platzman, P.~W.~Anderson,
\prb {\bf 19}, 5211 (1979).

\bibitem{Yoshioka}D.~Yoshioka and H.~Fukuyama,
J.\ Phys.\ Soc.\ Jpn.\ {\bf 47}, 394 (1979);
D.~Yoshioka and P.~A.~Lee,
\prb {\bf 27}, 4986 (1983).

\bibitem{Maki}K.~Maki and X. Zotos,
\prb {\bf 28}, 4349 (1983).

\bibitem{Laughlin}R.~B.~Laughlin,
\prl {\bf 50}, 1395 (1983).

\bibitem{Jain} J.~K.~Jain,
\prl {\bf 63}, 199 (1989).

\bibitem{HLR} B.~I.~Halperin, P.~A.~Lee, and N.~Read,
\prb {\bf 47}, 7312 (1993).

\bibitem{Sondhi} S.~L.~Sondhi, A.~Karlhede, S.~A.~Kivelson, and
E.~H.~Rezayi,
\prb {\bf 47}, 16~419 (1993).

\bibitem{Koulakov} A.~A.~Koulakov, M.~M.~Fogler, and B.~I.~Shklovskii,
\prl {\bf 76}, 499 (1996).

\bibitem{Belkhir_Morf} L.~Belkhir and J.~K.~Jain,
Solid\ State\ Commun.\ {\bf 94},
107 (1995); R.~Morf and N.~d'Ambrumenil,
\prl {\bf 74}, 5116 (1995).

\bibitem{Wu} X.-G.~Wu and S.~L.~Sondhi,
\prb {\bf 51}, 14~725 (1995).

\bibitem{Aleiner}I.~L.~Aleiner and L.~I.~Glazman,
\prb {\bf 52}, 11~296 (1995).

\bibitem{Enhancement} J.~F.~Janak,
Phys.\ Rev. {\bf 174}, 1416 (1969);
T.~Ando and Y.~Uemura,
J.\ Phys.\ Soc.\ Jpn. {\bf 35}, 1456 (1973).

\bibitem{Comment on exchange} Aleiner and Glazman neglected this term because
it is small in the limit $N \gg r_s^{-2} \gg 1$ they considered.

\bibitem{Experiment enhancement}
R.~J.~Nicholas, R.~J.~Haug, K.~v.~Klitzing, and G.~Weimann,
\prb {\bf 37}, 1294 (1988);
A.~Usher, R.~J.~Nicholas, J.~J.~Harris, and C.~T.~Foxton,
\prb {\bf 41}, 1129 (1990).

\bibitem{Kivelson} S.~Kivelson, C.~Kallin, D.~P.~Arovas,
and J.~R.~Schrieffer,
\prb {\bf 36}, 1620 (1987).

\bibitem{Seul} For review, see M.~Seul and D.~Andelman,
Science, {\bf 267}, 476 (1995).

\bibitem{Ashoori} R.~C.~Ashoori, J.~A.~Lebens, N.~P.~Bigelow,
and R.~H.~Silsbee,
\prl {\bf 64}, 681 (1990).

\bibitem{Eisenstein} J.~P.~Eisenstein, L.~N.~Pfeiffer, and K.~W.~West,
\prl {\bf 69}, 3804 (1992);
Surf.\ Sci.\ {\bf 305}, 393 (1994).

\bibitem{Turner} N. Turner, J. T. Nicholls, K. M. Brown, E. H. Linfield,
M. Pepper, D. A. Ritchie, and G. A. C. Jones, preprint cond-mat/9503040.

\bibitem{Aleiner_hydro} I.~L.~Aleiner, H.~U.~Baranger, and L.~I.~Glazman,
\prl {\bf 74}, 3435 (1995).

\bibitem{Levitov} L.~S.~Levitov and A.~V.~Shytov,
preprint cond-mat/9507058.

\bibitem{Efros} A.~L.~Efros,
Solid\ State\ Commun. {\bf 65}, 1281 (1988);
{\bf 67}, 1019 (1989);
{\bf 70}, 253 (1989).

\bibitem{Stormer} H.~L.~Stormer, K.~W.~Baldwin, L.~N.~Pfeiffer, and
K.~W.~West,
Solid\ State\ Commun.\ {\bf 84}, 95 (1992).

\bibitem{Chklovskii} D.~B.~Chklovskii, K.~A.~Matveev, and B.~I.~Shklovskii,
\prb {\bf 47}, 12~605 (1993).

\bibitem{Comment on peaks} The height of the peaks has the universal value of
$e^2 / 2 h$. To see this, one can use, e.g., the theory of Ref.~\cite{Ruzin}.
Actually, our theory justifies the main assumption made in this reference
that the separation of phases with $\nu_N = 0$ and $\nu_N = 1$ takes place.

\bibitem{Ruzin} I.~M.~Ruzin and A.~M.~Dykhne,
\prb {\bf 50}, 2369 (1994).

\bibitem{Gradshtein} I.~S.~Gradshtein and I.~M.~Ryzhik, {\it Tables of
Integrals, Series, and Products} (Academic Press, Boston, 1994).

\bibitem{MacDonald} A.~H.~MacDonald and G.~C.~Aers,
\prb {\bf 34}, 2906 (1986).

\bibitem{Kukushkin} I.~V.~Kukushkin, S.~V.~Meshkov, and V.~B.~Timofeev,
Usp.\ Fiz.\ Nauk {\bf 155}, 219 (1988)
[Sov.\ Phys.\ Usp. {\bf 31}, 511 (1988)].

\bibitem{Chamon} C.~de~C.~Chamon and X.~G.~Wen,
\prb {\bf 49}, 8227 (1994).

\bibitem{Comment on order} To determine the order of the transition,
we have considered an alternative (second-order transition) scenario, which
is as follows. First, at certain $\bar\nu_N > \nu_N^\ast$, the width of the
stripes acquires a periodic modulation in $y$. As $\bar\nu_N$ decreases, the
amplitude of the modulation increases. Eventually, at $\bar\nu_N =
\nu_N^\ast$, the stripes break into isolated ``bubbles''. We have found,
however, that the decribed above modulation first arises at $\bar\nu_N =
0.375$, which is smaller that $\bar\nu_N = 0.39$ where the first-order
transition occurs.  Thus, the proposed above mechanism is not realized, and
the transition is of the first order.

\bibitem{Comment on many-body} Strictly speaking, due to a nonzero overlap
the energy of the system is not reduced to the sum of solely two-particle
interaction energies. However, this appears to be a small effect~\cite{Maki}.

\bibitem{Efros_Shklovskii}B.~I.~Shklovskii and A.~L.~Efros,
{\em Electronic Properties of Doped Semiconductors} (Springer, New York, 1984).


\bibitem{Zheng}L.~Zheng and A.~H.~MacDonald,
\prb {\bf 47}, 10~619 (1993).

\bibitem{Shrieffer}J.~R.~Shrieffer, D.~J.~Scalapino, and J.~W.~Wilkins,
\prl {\bf 10}, 336 (1963).

\bibitem{Murphy}S.~Q~Murphy, J.~P.~Eisenstein, L.~N.~Pfeiffer, and K.~W.~West,
\prb {\bf 52}, 14~825 (1995).

\bibitem{Glozman} I.~Glozman, C.~E.~Johnson, and H.-W.~Jiang,
\prl {\bf 74}, 594 (1995).

\bibitem{Buks} E.~Buks, M.~Heiblum, and Hadas Shtrikman,
\prb {\bf 49}, 14 790 (1994);
E.~Buks, M.~Heiblum, Y.~Levinson, and Hadas Shtrikman,
Semicond.\ Sci.\ Technol.\ {\bf 9}, 2031 (1994).

\bibitem{Comment on tails} The tails of the tunneling peak in
both cases are neither Lorentzian nor Gaussian but rather very nearly
exponential: $\ln[G(V) / G(0)] \sim -e V / \gamma_0$.  Here $\gamma_0 \sim U^2
\tau / \hbar$. This asymptotical behavior holds at $e V \gg \gamma_0$ in the
``quantum'' case $n \gg n_i$ and at $e V \gg \frac{\hbar}{\tau}$ in the
``quasiclassical'' case $n \ll n_i$. Thus, at the crossover point $n \sim
n_i$, the exponential dependence is to be observed in the entire range of
bias voltages.

\bibitem{Comment on SdH}In the ``quantum'' case, $n \gg n_i$, the
Shubnikov-de Haas oscillations are described by the Dingle formula,
$\delta\rho_{\rm xx} / {\rho_{\rm xx}^0} \propto -{\rm
e}^{-{\pi}/{\omega_c\tau}}\cos(\pi\nu)$, which applies when $\omega_c\tau
\lesssim \pi$. (The temperature is assumed to be low enough).  In the
``quasiclassical'' case, $n \ll n_i$, the QHE effect gives way to
Shubnikov-de Haas oscillations at $\hbar\omega_c \lesssim \pi U$. It is
interesting to note that the conventional Dingle law holds only at much
weaker magnetic fields, $\hbar\omega_c \ll (\hbar v_{\rm F} / d)$.  In the
intermediate range, $(\hbar v_{\rm F} / d) \ll \hbar\omega_c \ll U$, the
Dingle law is replaced by a different formula~\cite{Mirlin}:
$\delta{\rho_{\rm xx}} / {\rho_{\rm xx}} \propto -{\rm e}^{-{2 \pi^2
U^2}/{\hbar^2\omega_c^2}}\cos(\pi\nu)$.  Recently, this dependence has been
indeed found in the experiment (P.~T.~Coleridge, private communication).

\bibitem{Mirlin} A.~G.~Aronov, E.~Altshuler, A.~D.~Mirlin, and P.~W\"offle,
\prb {\bf 52}, 4708 (1995);
A.~D.~Mirlin, E.~Altshuler, and P.~W\"olfle,
preprint cond-mat/9507081.

\bibitem{Fogler} M.~M.~Fogler and B.~I.~Shklovskii,
\prb {\bf 52}, 17~366 (1995).

\bibitem{Raikh} M.~E.~Raikh and T.~V.~Shahbazyan,
\prb {\bf 47}, 1522 (1993).

\bibitem{Comment on shape} If $l \ll d$, the theory of Raikh and
Shahbazyan~\cite{Raikh} applies. The LLs have a Gaussian form [similar to
Eq.~(\ref{A Gauss})]. In the opposite case, $l \gg d$, the SCBA (Self
Consistent Born Approximation) is valid and the LLs have a semielliptic form
(see Ref.~\onlinecite{Raikh} and references therein).

\bibitem{Yang} S.-R. Eric Yang and A.~H.~MacDonald,
\prl {\bf 70}, 4110 (1993).

\bibitem{Eisenstein_95} In strong magnetic fields, this effect
has been experimentally studied by J.~P.~Eisenstein, L.~N.~Pffeifer, and
K.~W.~West,
\prl {\bf 74}, 1419 (1995).

\bibitem{Rokhinson} L.~P.~Rokhinson, B.~Su, and V.~J.~Goldman,
Solid\ State\ Commun.\ {\bf 96}, 309 (1995);
L.~P.~Rokhinson and V.~J.~Goldman,
unpublished.

\bibitem{Dyson}See P.~M.~Bleher, Z.~Cheng, F.~J.~Dyson, and J.~L.~Lebowitz,
Commun.\ Math.\ Phys.\ {\bf 154}, 433 (1993).

\end{references}
\end{document}